\documentclass[fleqn,usenatbib]{mnras}
\hypersetup{draft}

\usepackage[T1]{fontenc}
\usepackage{ae,aecompl}

\usepackage{graphicx}	
\usepackage{amsmath}	
\usepackage{amssymb}
\usepackage{newtxtext,newtxmath}

\title[Triple evolution and galactic tides]{Chaotic dynamics of wide triples induced by galactic tides:\\ a novel channel for producing compact binaries, mergers and collisions}

\author[Grishin and Perets]{
Evgeni Grishin,$^{1,2,3}$\thanks{E-mail: evgeni.grishin@monash.edu (EG)}
Hagai B. Perets,$^{1}$
\\
$^{1}$Physics Department, Technion - Israel Institute of Technology, 3200003, Haifa, Israel\\
$^{2}$School of Physics and Astronomy, Monash University, Victoria, Australia, Clayton, Victoria 3800, Australia\\
$^{3}$The ARC Centre of Excellence for Gravitational Wave Discovery - OzGrav, Australia\\
}

\date{Accepted XXX. Received YYY; in original form ZZZ}

\pubyear{2021}

\begin{document}
\label{firstpage}
\pagerange{\pageref{firstpage}--\pageref{lastpage}}
\maketitle

\begin{abstract}
Recent surveys show that wide ($>10^4$ AU) binaries and triples are abundant in the field. We study the long-term evolution of wide hierarchical triple systems and the role played by the Galactic tidal (GT) field. We find that when the timescales of the secular von-Ziepel-Lidov-Kozai and the GT oscillations are comparable, triple evolution becomes chaotic which leads to extreme eccentricities. Consequently, the close pericentre approaches of the inner-binary components lead to strong interactions, mergers and collisions. We use a novel secular evolution code to quantify the key parameters and carry out a population-synthesis study of low and intermediate-mass wide-orbit triples. We find that in $\sim9\%$ of low-mass wide-triples the inner main-sequence binaries collide or tidally-inspiral within $10\ \rm Gyr$, with direct collisions are $6$ times more likely to occur.
For the intermediate-mass sample, $\sim7.6\%$ of the systems merge or inspiral with roughly equal probabilities. We discuss the relative fractions of different stellar merger/inspiral outcomes as a function of their evolutionary stage (Main-Sequence, MS; Red-Giant, RG; or White-Dwarf, WD), their transient electromagnetic signatures and the final products of the merger/inspiral. In particular, the rate of WD-WD direct-collisions that lead to type-Ia Supernovae is comparable to other dynamical channels and accounts for at most $0.1\%$ of the observed rate. RG inspirals provide a novel channel for the formation of eccentric common-envelope-evolution binaries. The catalysis of mergers/collisions in triples due to GT could explain a significant fraction, or even the vast majority, of blue-stragglers in the field, produce progenitors for cataclysmic-variables, and give-rise to mergers/collisions of double-RG binaries.
\end{abstract}




\section{Introduction} \label{sec:1}
The close interactions between stars through mass transfer, common-envelope evolution (CEE), mergers and collsions play an important part in the evolution of isolated close interacting binary systems. 
In triple systems, secular perturbations, such as Von Ziepel-Lidov-Kozai oscillations \citep[ZLK;][]{vZ1910, lid62, koz62} and quasi-secular evolution \citep{ap12,luo16, gpf18} induced by the outer third companion, can drive the inner binaries into highly eccentric configurations. The close pericentre approach of such binaries then gives rise to strongely interacting binaries. 
Indeed, triple secular and quasi-secular dynamics ware proposed as catalyze the formation of a wide variety of exotic stars and binaries, as well as various explosive and transient events, e.g. blue stragglers \citep{per_fab}, X-ray sources \citep{nao+16}, gravitational-wave (GW) mergers \citep{ap12,ant2014, antog2014, ant2017, ll18, hoang2018, martinez2020, tbh, vg2021}  and type Ia SNe \citep{katz2011,tho11}, however, the latter can only produce a fraction of the inferred typa Ia SN rate \citep{toonen2018}, (see \citealp{naoz_review} for a review on many of these issues). Such evolution becomes particularly important under appropriate conditions in which the relative inclinations between the inner binary and the outer binary in a triple are high.  

 Secular evolution of stellar-binaries could also occur due to perturbation by external non-stellar potentials, such as binaries perturbed by a massive black hole in galactic nuclei \citep{ap12,pro+15}; binaries in cluster potentials \citep{hamilton1, hamilton2, hamilton3}, or wide binaries perturbed by the Galactic tidal field \citep{ccg17}. In fact, already decades ago the secular evolution of Oort cloud objects due to the Galactic tidal field was shown to be important for the production of highly eccentric comets and the formation of long-period and sun-grazing comets \cite[e.g][]{ht86}. The timescale for such oscillations, however, is long, and of order of $\rm Gyr$ for Oort cloud objects in the Solar neighbourhood. 
 
 Secular triple evolution becomes even more complex when additional outer perturbations. \protect{\citep{hamers15}} considered the secular evolution of quadruple hierarchical systems. In nuclear star clusters, a binary around a supermassive black hole can undergo chaotic evolution if the cluster has non-spherical potential \protect{\citep{petrovich17, bub2020}}, or if additional stochastic torque is present due to resonant relaxaion \protect{\cite{hamers_vrr}}. Indeed, the sensitivity of triple evolution to the mutual inclinations between the inner and outer binary makes them even more suceptible to significant secular evolution by an external potential, which could also arise from a non-stellar external perturbation. Here we study, for the first time, the coupling of the secular evolution of wide triple systems with the external Galactic tidal field and show it has major implplications for the formation of closely interacting binaries in the field.
 
 The secular evolution due to the galactic tidal field is very similar to ZLK oscillations, with small quantitative variations underlying the same mechanism \citep{hamilton1, hamilton2, hamilton3}. 
  Hence the systems of wide triple and galactic tides are qualitatively similar to a quadruple hierarchial system. These systems are known to experience chaotic dynamics if the secular frequencies are comparable \citep{hamers15, hl17, glp18, hamers19}. Similar ideas of chaotic dynamics had been carried out for chaotic evolution of stellar spin \citep{sl14, sl15} and the spin of binary black holes in the final moments of their merger \citep{ll18}, and they all rely on the idea of overlapping resonances \citep{chirikov79}.

Tidal effects coupling to the secular evolution of triples become important for very wide systems, (typically $>10^4$ AU). Such systems are not rare. Half of Solar-type stars and a quarter of lower-mass M-dwarf stars are are part of binary systems \citep{raghavan2010, duchene2013}, while multiple stars are even more common for higher masses. The Gaia mission \citep{gaia_mission} had revolutionized modern astrometry and provided unprecidented data of more than a billions stars in several data releases \citep{gaia-dr1, gaia-dr2, gaia-dr3}. Such surveys enabled the identification of wide binary systems \citep{el-badry2019,el-badry-twin,hartman2020,el-badry2021}. As wide binaries are generally detached, they do not affect each other through the major course of their evolution.

Wide binaries can serve as a tool to study galactic dynamics \citep{weinberg1987, tremaine2010}, the galactic tidal field and the exitence of MACHO dark matter  \citep{macho_dm}. They are used to constrain stellar flarings \citep{2morgan2016}, natal kicks of white dwarfs (WDs) \citep{el-bardy2018} and neutron stars (NS) \citep{nk}, and CEE timescales \citep{ce1, ce2}. Moreover, wide binaries and triple are suceptible to collisional dynamics in the field and the production of closely interacting binaries, merger products,  X-ray sources and GWs sources \citep{kaib2014,michaely2016,michaely2019,michaely2021,mic_sha21}. Recent observations report a few percents of stellar systems are in ultra-wide binaries \citep{hwa+21}, and given the high fraction of close binaries \cite{raghavan2010}, about half of the ultra-wide binaries whould be ultra-wide triples, while triple configurations at larger masses are even more common \citep{nk}. 
 
In this paper we study the chaotic dynamics of wide triples induced by the galactic tide. Without the galactic tide, the range of initial conditions that lead to highly eccentric encounter and significant close interaction, is narrow (namely due to the stringent constrain on the mutual inclination). We outline the parameter space most feasible for chaotic evolution, and develop the secular code \href{https://github.com/eugeneg88/SecuLab}{\texttt{SecuLab}} to probe its complex evolution, which includes various relevant physical processes as described in detail below. We showcase the importance of this wide chaotic-triple-galactic-tide-evolution (CATGATE) channel and its contribution to the overall rates of the aforementioned transient phenomena and newly formed star and stellar remnants in the Galaxy, and in other galaxies.

This paper is organized as follows. Sec. \ref{sec:2} describes the various details of the secular evolution itself. We review some aspects of LK evolution in \ref{sec:2.1}, galactic tides in \ref{sec:2.2}, consevative forces induced by rotation, tides, and general relativity (GR) in \ref{sec:2.3}, as well as dissipative forces, and outline a transition point from equilibrium to dynamical tides in \ref{sec:2.4}. Sec. \ref{sec:3} describes the evolution of particular chaotic systems and qualitative trends. In \ref{sec:3.1} we showcase the chaotic evolution of individual systems, in \ref{sec:3.2} we outline the choice of our initial and stopping conditios, and our simplified modelling of stellar evolution in \ref{sec:3.3}. Sec. \ref{sec:4} presents the results from a population synthesis study. We present results for low mass stars in \ref{sec:4.1} and for massive stars up to $8M_\odot$ in \ref{sec:4.2}. In sec. \ref{sec:5} we interpret and discuss our results in context of our transients universe and the formation channels of various stars and stellar remnants. We discuss overall rates in \ref{5.1} and astrophysical implications in \ref{implications}. We also revisit our assumptions and caveats in sec. \ref{sec. 5.4}. Finally sec. \ref{sec:6} summarizes the main findings of this paper.

\section{Secular dynamics} \label{sec:2}
In this section we overview the different processes that govern the secular dynamics of triple systems (\citealp[see][]{naoz_review} for a review), as well as additional forces due external and internal perturbations. In addition, we derive an elegant form of the Galactic tide Hamiltonian and subsequent secular equations and stress their similarities to ZLK cycles (cf. sec. \ref{sec:2.2}).

\subsection{von Ziepel-Lidov-Kozai oscillations} \label{sec:2.1}
Consider a hierarchial triple system with an inner binary of masses $m_0$ and $m_1$, separation $a_1$ and eccentricity $e_1$, perturbed by a distant companion of mass $m_2$ at separation $a_2 \gg a_1$ and eccentricity $e_2$. The interaction term is expanded in multipoles of increasing powers of $a_1/a_2$ and double-averaged over the two fast mean motion angles of both orbits. The resulting secular Hamiltonian is responsible for the secular evolution of the system. 

If the relative inclination between the orbital planes, $i$ is sufficiently large, then coherent oscillations could bring the inner binary to large eccentricities  \citep{vZ1910, lid62, koz62}. The typical secular time in the quadrupole approximation (i.e. truncation up to order $(a_1/a_2)^2$ is \footnote{The actual timescale depends on the inclination. \cite{ant15} derived the timescale more rigorously and found a correction factor of $16/15$.}

\begin{equation}
    \tau_{{\rm sec}}=\frac{1}{2\pi}\frac{m_{\rm tot}}{m_2}\frac{P_{2}^{2}}{P_{1}}(1-e_2^{2})^{3/2}, \label{eq:tsec_in}
\end{equation}
where $P_1 = 2\pi (Gm_{\rm in}/a_1^3)^{-1/2}$ and $P_2 = 2\pi (Gm_{\rm tot}/a_2^3)^{-1/2}$ are the inner and outer binary periods, respectively, with $m_{\rm in} = m_0 + m_1$ is the inner binary mass and $m_{\rm tot} = m_{\rm in} + m_2$ is the total mass.

In the test particle limit (i.e. where the outer binary angular momentum dominates) the maximal eccentricity depends only on the initial mutual inclination\footnote{It is true for a orbit where the argument of pericentre, $\omega$, circulates. For librating orbits the eccentricity could more constrained if the initial conditions are close to the fixed point.}. And the existense of a LK resonance is restricted to inclinations of $|\cos i| \le\sqrt{3/5}$.

When the orbit is not sufficiently hierarchical, the averaging of the orbit of the tertiary may no longer be accurate. \cite{ap12} pointed out the key importance of quasi-secular regime for triple evolution, where the long-term orbital evolution can be altered due to short term corrections.  Corrections to the double-averaged secular Hamiltonian occur on the 'single-averaged' (SA)
timescale, and their strength is characterized by the dimensionless parameter \citep{luo16},
\begin{equation}
\epsilon_{\rm SA} = \frac{P_2}{2\pi \tau_{\rm sec}} = \left(\frac{a_1}{a_{2}(1 - e_{2}^2)}\right)^{3/2}\left(\frac{m_2^2}{(m_{\rm in} + m_2)m_{\rm in}}\right)^{1/2}. \label{eq:eps_sa}
\end{equation}
\cite{gpf18} obtained analytical modified expressions for $e_{\rm max}$ and the critical inclination for ZLK resonance. The astrophysical implications of the correction for double-averaging are also relevant for evection resonances and stability of irregular satellites \citep{grish17}, formation of contact binaries in the Kuiper belt \citep{gri20, roz2020}, Hot Jupiters and gravitational wave mergers of binary black holes \citep{ap12, ll18, gpf18, fg19}  and type Ia supernovae \citep{katz2011, hk18, toonen2018}.

Although for the very wide stellar triples we consider, we expect $\epsilon_{\rm SA}$ to be small, due to the large $a_2$ values, the secular evolution is not affected much. However, short term fluctuations could render the eccentricity to be unconstrained, provided that $\sqrt{1-e^2} \lesssim \epsilon_{\rm SA}^2$, which potentially occurs in our case. We defer the analysis of these non-secular effects to sec. \ref{sec. 5.4}.

\subsubsection{Quadruple systems and chaos} \label{2.1.1}
Quadruple hierarchical system can be divided between two type of architectures, either $2+2$ quadruple system, where two inner binaries perturb each other, or $(2+1)+1$ binaries, where the heirarchy is nested \citep{hamers15, hl17}.

Consider a fourth body of mass $m_3$ and separation $a_3 \gg a_2$ and eccentricity $e_3$ around the centre of mass of the hierarchical triple, such that the configuration is a $3+1$ quadruple system. In this case there are three binaries, with inner, intermediate and outer binary denoted by A, B and C, respectively. The the intermediate binary B is then perturbed on a secular timescale $\tau_{\rm sec, out}$ which is given by Eq. (\ref{eq:tsec_in}, where the inner binary period is $P_2$ and the outer binary period is $P_3 = 2\pi (G m_{\rm tot}/a_3)^{-1/2}$, with $m_{\rm tot} = m_0 + m_1 + m_2 + m_3$.

The qualitative behavior is determined by the 'adiabatic parameter'  $\mathcal{R}_0 = \tau_{\rm sec} / \tau_{\rm sec, out}$ \citep{sl14, sl15, hamers15, glp18}. If $\mathcal{R}_0 \ll 1$, the angular momentum of the innermost binary A  precesses around the angular momentum of the intermediate binary B, which in turn slowly precess around the angular momentum of binary C. If $\mathcal{R}_0 \gg 1$, the angular momentum of both binary A and B  precess around the angular momentum of the intermediate binary C, where the binaries A and B are effectively decoupled. 

In the case where $\mathcal{R}_0 \sim 1$, the evolution is complex and chaotic due to resonanse overlap of both secular frequencies \citep{chirikov79}. \cite{sl14, sl15} have analyzed analytically a related problem of the evolution of the stellar spin orbited by an eccentric Jovian planet in a binary star system, while \cite{glp18} connected and extended the analytical theory to hierarchical $3+1$ quadruple systems, mapping out the phase-space of chaotic evolution. Recently, \cite{hamers19} compared the results of a population synthesis of quadruple main sequence stars to the observed properties and concluded that the results fit reasonably well for $3+1$ systems, but not for $2+2$ systems.

\subsection{Galactic tides} \label{sec:2.2}
When  a binary is on a wide orbit ($a > 10^4 \rm au$), then is can be significantly influenced by the tidal field of the Galactic potential.  \cite{ht86} first studied the effect of\ Galactic tide on the orbits or Oort-cloud comets and found it to be the main effect that drives comets onto Sun-grazing orbits; the perturbations from Galactic tide accumulate similarly to ZLK oscillations, enhancing the eccentricity of the comet until it grazes the Sun.

Here we follow the simplified model of \cite{ht86}, taking into account only the leading term of the vertical tide, neglecting perturbations on other directions (which are about $\sim 15$ times smaller, \citealp{ht86}). Various studies looked on the additional terms \citep[e.g][]{veras13, ccg17} and found the behaviour to be qualitatively similar to \cite{ht86}. 

The potantial due to Galactic tide in the rotating frame centered on the midplane at distance $R = 8 \ \rm kpc$ from the Galactic centre is  
\begin{equation}
    U(x,y,z) =-\frac{Gm_{\rm tot}}{\sqrt{x^2+y^2+z^2}}+2\pi G\rho_{0}z^{2}, \label{eq:gt}
\end{equation}
where $m_{\rm tot}$ is the total mass of the binary and $\rho_0 = 0.185\ \rm M_{\odot} \ pc^{-3} $ is the local density at the solar neighbourhood. 
After performing secular averaging on the mean anomaly, the resulting Hamiltonian is:
\begin{equation}
    H_{{\rm GT}}=-\frac{Gm_{{\rm tot}}}{2a}+\pi G \rho_{0}a^{2}\sin^{2}i(1-e^{2}+5e^{2}\sin^{2}\omega), \label{eq:gtsec}
\end{equation}
where $i$ is the inclination angle between the binary and the Galactic disc, and $\omega$ is the argument of pericentre. Expressing the Hamiltonian in terms of the canonical Delaunay elements and writing down the equations of motion, the only difference is in the evolution for $\omega$:

\begin{equation}
\frac{d\omega}{dt} = \frac{3}{4 \tau_{\rm sec}}\left[ (\gamma+1)(1-e^{2})+5\sin^{2}\omega\left(e^{2}-\sin^{2}i\right)\right], 
 \label{eq:domega_dt}   
\end{equation}
 where $\gamma = 1$ for ZLK cycles and $\gamma = 0$ for Galactic tides.
 
 It is then straightforward to show than the maximal eccentricity is 

\begin{equation}
    e_{\rm max} = \sqrt{1 - \frac{5}{4 - \gamma }\cos^2{i}} \label{eq:emax2}
\end{equation}
for a critical inclination of $\cos^{2}i_c<(4-\gamma)/5$. The secular timescale is 
\begin{equation}
    \tau_{\rm GT} = \frac{3}{8\pi G\rho_0} \sqrt{ \frac{Gm_{\rm tot}}{a^3} } = 0.89 \left( \frac{m_{\rm tot}}{M_\odot} \right)^{1/2}  \left( \frac{a}{10^4 {\rm AU}} \right)^{-3/2} \rm Gyr  \label{eq:tsecgt}
\end{equation}
When the secular timescales $\tau_{\rm sec}$ and $\tau_{\rm GT}$ (Eq. \ref{eq:tsec_in} and \ref{eq:tsecgt}) are comparable, the dynamics will be chaotic, similar to the dynamics of hierarchical quadruples discussed in sec. \ref{2.1.1}, and will be demonstrated in sec. \ref{sec:3}.

\subsection{Conservative forces and extra precession} \label{sec:2.3}

The ZLK mechanism requires that $\omega$ will slowly evolve, and a resonance will occur if, on average, $\dot{\omega} = 0$. In reality, $\omega$ is also precessing due to General-Relativistic (GR) corrections, or tidal and rotational bulges that break the spherical symmetry. If the precession is too fast compared with the ZLK secular timescale, the ZLK oscillations will be quenched. 

The maximal eccentricity can be limited even for slow additional precession. \cite{lml15} found an analytical expression for the maximal eccentricity:

\begin{align}
\epsilon_{\rm GR} \left( \frac{1}{j_{\rm min}} - 1 \right)  + \frac{\epsilon_{{\rm Tide}}}{15}\left(\frac{f_1(e_{{\rm max}})}{8j_{{\rm min}}^{9}}-1\right) & =\frac{9e_{{\rm max}}^{2}}{8j_{{\rm min}}^{2}}\left(j_{{\rm min}}^{2}-\frac{5}{3}\cos^{2}i\right). \label{etide}
\end{align}
where  $f_1(e) = 1 + 3e^2 + 3e^4/8$, $j_{\rm min} = \sqrt{1 - e_{\rm max}^2}$ and

\begin{align} 
\epsilon_{{\rm GR}}& \equiv \frac{3m_{{\rm bin}}(1-e_{2}^{2})^{3/2}}{m_{2}}\left(\frac{a_{2}}{a_1}\right)^{3}\frac{r_{g}}{a_1}\nonumber \\
\epsilon_{{\rm Tide}} & \equiv\frac{15m_{0}^{2}a_{2}^{3}(1-e_{2}^{2})^{3/2}(2k_1)R_{1}^{5}}{a_1^{8}m_{1}m_{2}}, \label{eq:epstide-1}
\end{align}
are the relative precession rates compared to ZLK precession rate. In Eq. (\ref{eq:epstide-1}), $r_g \equiv G m_{\rm bin} /c^2$ is the gravitational radius, $k_1$ is the apsidal motion constant and $R_1$ is the radius of body 1. The maximal eccentricity is small for $\epsilon_{\rm GR/Tide} \gg 1$ and reduces to the standard ZLK maximal eccentricity for $\epsilon_{\rm GR/Tide} \to 0$ (cf. \citealp{lml15} for details).

Since the dependence on tides is extremely sensitive to the initial inner separation we can approximately solve for Eq. \ref{etide} by taking either GR or tidal terms under the assumptions of   $\cos i_0=0$ and $j_{\rm min}=(1-e^2_{\rm max})^{0.5}\ll1$. For GR or tides only, the solutions are 
\begin{equation}
    j_{\rm GR} = \frac{8\epsilon_{\rm GR}}{9};\quad \quad j_{\rm Tide} = \left(\frac{7\epsilon_{\rm Tide}}{216}\right)^{1/9}.\label{eq:j_approx}
\end{equation}
If $j$ from Eq. \ref{eq:j_approx} is $>1$, then $e_{\rm max}=0$. 

\begin{figure}
    \centering
    \includegraphics[width=8.5cm]{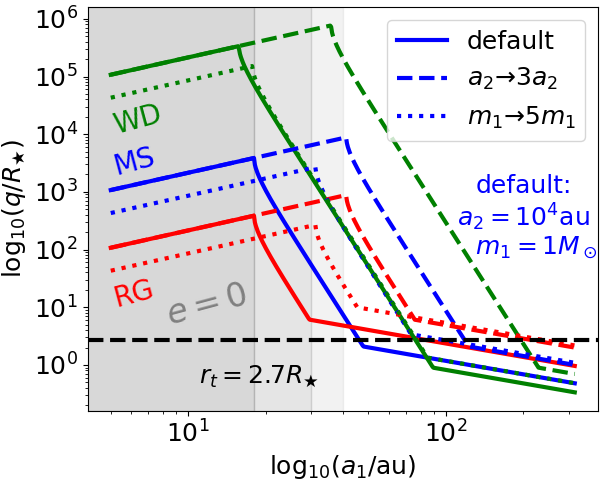}
    \caption{The minimal pericentre possible for the inner binary due to secular evolution. The other masses are $m_2=m_3=1M_\odot$. The radius is star 1 is $R_\star = R_\odot(m_1/M_\odot)^{0.57}$ for MS stars. The final mass of the WD is given by Eq. \ref{eq:mwd}. For RGs, the radius is 10 times larger, and for WDs it is $0.01R_\odot$.  The apsidal motion constant is $k_1=0.014$ for MS, and $0.1$ for RGs and WDs. Blue curves indicate MS stars, red curves indicate RG stars and green curves indicate WDs. Solid lines are for the default model of $a_2=10^4\ \rm au$ and $m_1=M_\odot$. Dashed lines are for increasing $a_2$ to $3\cdot 10^4\ \rm au$ and dotted lines are for increasing $m_1$ to $5M_\odot$ (while keeping $a_2$ fixed at $10^4\ \rm au$). The horizontal dashed black line indicates the proxy for the tidal raidus at $2.7R_\odot$. The grey areas indicate regions where $\epsilon_{\rm GR}\gg1$ and hence the eccentricity is significantly suppressed at zero. }
    \label{fig:min_p}
\end{figure}

Fig. \ref{fig:min_p} shows the minimal pericentre (which is derived from the maximal eccentricity) from solving Eq. \ref{etide} approximately, using $e_{\rm max}=(1-j^2)^{1/2}$ where $j=\max(j_{\rm GR}, j_{\rm Tide})$. The orbits of both binaries are circular and the inclination is $\cos i_0=0$. Each colour indicates a different stellar type of star $m_1$, while the other stars are assumed to be point Solar mass stars. The other parameters are described in the caption.

We see that there are three typical phases: 

i) The eccentricity is essentially zero due to large $\epsilon_{\rm GR}$, and the pericentre is just $a_1$ and by definition increases with $a_1$. This occurs up to separations of $a_1 \approx 20-40\ \rm au$. 

ii) The minimal pericentre is rapidly decreasing with $a_1$. This occurs since $\epsilon_{\rm Tide} \ll \epsilon_{\rm GR} \ll 1$ in Eq. \ref{eq:j_approx} (recall that $\epsilon_{\rm GR} \propto a_1^{-4}$), and stops roughly at $a_1 \sim 50\ \rm au$ for the default MS model (blue solid line), close to the tidal disruption limit. For other models the range is from $30-200 \rm au$.

iii) The pericentre is slowly decreasing with $a_1$. This occurs when $\epsilon_{\rm GR}<\epsilon_{\rm Tide} \ll 1$, and $j$ depends on $\epsilon_{\rm GR}$ only as a power of $1/9$. Most of the models are below the tidal disruption limit at this point and occur around $a_1\sim 50-100\ \rm au$.  

To summarize, on average, for ultra-wide triples (when $a_2 \gtrsim 10^4\ \rm au$) extra precession completly quenches ZLK oscillation for $a_1\lesssim 20\ \rm au$ and prevents collisions and disruptions for $a_1 \lesssim 50-100\ \rm au$. Triples with $a_1 \gtrsim 100\ \rm au$ could potentially have unconstrained eccentricities that even allow direct collisions of the inner binary.

\subsection{Dissipative forces} \label{sec:2.4}
When the innermost two bodies are close enough, the dissipation timescale can be short enough to affect the dynamics. The dissipative forces usually scale sharply with the instantaneous separation, and therefore often approximated by impulsive dissipation at the pericentre for highly eccentric orbits. For compact objects, dissipation from gravitational wave emission is the dominant contribution, while for planets, MS and RG stars, tidal dissipation is more important \citep{FT07, p15_1}.

\subsubsection{Equilibrium tides}
 \cite{hut81} studied the equilibrium tide model, where a constant tidal bulge  lags behind the line connecting the two bodies, and creates a torque that changes the angular momentum, which in turn drives the system into synchronous state.
 
 The synchronisation time is an order of magnitude shorter than the circularisation time. Therefore the system spends most of its time in a pseudo-synchronous state, where the spin of the body is aligned with the orbit, and the spin rate is 
 \begin{equation}
\Omega_{\rm ps}(e) = \frac{2 \pi}{P_1} \frac{f_2(e)}{(1-e^2)^{3/2}f_1(e)} \label{eq:ps}
 \end{equation}

Under this approximation, the change in the semi-major axis and eccentricity vector is 
\begin{align}
    \frac{da_1}{dt} = & - \frac{2}{9t_{\rm TF}}\frac{a_1}{(1-e_1^2)^{15/2}} \left( f_1(e_1) - (1-e_1^2)^{3/2}f_2(e_1)\frac{\Omega_{\rm ps} (e_1)}{(2 \pi / P_1)} \right) \nonumber \\
    \frac{d\boldsymbol{e}_1}{dt} =  & -\frac{1}{t_{\rm TF}}\frac{\boldsymbol{e}_1}{(1-e_1^2)^{13/2}}\left(f_3(e_1) -  (1-e_1^2)^{3/2}f_4(e_1)\frac{11 \Omega_{\rm ps} (e_1)}{18 (2 \pi / P_1)}\right)
    \label{eq:dedt_tide}
\end{align}
where the tidal friction time is \citep{FT07, p15_1}
\begin{equation}
    t_{\rm TF} = \frac{t_{\nu}}{9(1+2k_1)^2}\left( \frac{a_1}{R_1}\right)^8 \frac{m_1^2}{(m_1+m_0)m_0}, \label{eq:tf}
\end{equation}
here $t_{\nu}$ is the typical viscous time of body 1 and $k_1$ is the apsidal motion constant. Typical viscous times are around $\sim 5\  \rm yr$ for MS stars \citep[e.g. ][]{hamers21}. Under the pseudo-synchronisation aproximation, the angular momentum is conserved, and the dissipation of the energy is governed by the change of the eccentricity under the assumption of constant angular momentum.

The polynomial functions $f_i$ are (e.g. Eq. 13 in \citealp{mk18})
\begin{align}
    f_1(e)=&1+\frac{31e^2}{2} + \frac{255e^4}{8} + \frac{185e^6}{16} + \frac{25e^8}{64} \nonumber \\
    f_2(e)=&1+\frac{15e^2}{2} + \frac{45e^4}{8} + \frac{5e^6}{16} \nonumber \\
    f_3(e)=&1+\frac{15e^2}{2} + \frac{15e^4}{8} + \frac{5e^6}{64} \nonumber \\
    f_4(e)=&1+\frac{3e^2}{2} + \frac{1e^4}{8} \label{eq:fi}
\end{align}
\subsubsection{Dynamical tides}
The equilibrium tide model is valid for static stars, that usually have low eccentricity. Conversely, for highly eccentric or unbound binaries, the tidal bulge is raised essentially during the closest approach, and the subsequent energy is dissipated via non radial dynamical oscillations \citep{pt77, mardling1, mardling2, lai1997}, but the general trend is to drive the system into short period and circular orbit faster than in the equilibrium tide model. 

In a recent paper, \cite{mk18} included a simplified prescription for dynamical tides. Ignoring the chaotic phase, the energy dissipation in the leading terms of the dynamical modes is 
\begin{equation}
    \Delta E = f_{\rm dyn} \frac{m_0+m_1}{m_1} \frac{G m_0^2}{R_1} \left( \frac{R_1}{a_1 (1-e_1)} \right)^9
\end{equation}
which ralates to the total change in the semi-major axis:
\begin{equation}
    \frac{da_1}{dt}=-\frac{2a_1^2}{P_1} \frac{\Delta E}{G m_0 m_1}.
\end{equation}
Here $f_{\rm dyn}$ parametrizes the efficiency of dynamical tides, and varies from $f_{\rm dyn} \sim 0.03 -1$, depending on the properties stellar structure \citep{mk18}.

Note that in the discussion above the raised only on body 1, while the primary is effectively treated as a point mass. The roles can be reversed, and in reality both bodies experience tidal dissipation.

\subsubsection{Transitional eccentricity}

\begin{figure}
	\centering
	\includegraphics[width=8.2cm]{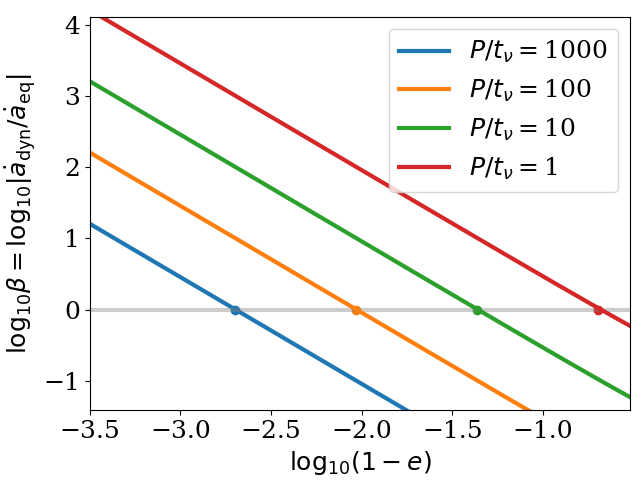}
		\caption{The ratio between the dissipation rate of dynamical and equilibrium tides as a function of the eccentricity. For each curves the parameters are $k_1 0.014, = f_{\rm dyn} =0.03$. The orbital periods are $P_1/t_{\nu} = 1000, 100, 10, 1$, which correspond to blue, orange, green and red lines, respectively. The thick dots are approximate solutions, dereived in Eq. (\ref{eq:e_trans}).}
	\label{fig2}
\end{figure}

Since equilibrium tides operate on almost circular binaries, while dynamical tides operate mostly on highly eccentric orbits, there is a typical value of a transitional eccentricity $e_{\rm trans}$ where both terms are equal. \cite{mk18} used $e_{\rm trans}=0.8$, somewhat similar value to \cite{wu18}, and also assumed that only one of the tidal prescriptions is valid. However, by doing so, the tidal dissipation is inherently discontinuous. 

Moreover, the transitional eccentricty is extremely sensitive to the assumed viscous time and the initial period of the binary. Comparing the dissipation rate for both prescriptions, the ratio between the energy dissipation rate is \citep[e.g.][]{rozner2021, glanz_roz21}
\begin{equation}
    \beta(e) \equiv \left| \frac{\dot{a}_{\rm dyn}}{\dot{a}_{\rm eq}} \right| = \frac{2f_{\rm dyn} R_1^3}{21Gm_1k_1\tau_L P_1} \frac{(1-e^2)^{15/2}}{e^2(1-e)^9 g(e)} \label{eq:beta}
\end{equation}
where $\tau_L$ is the lag time and 
\begin{align}
g(e) = \frac{1 + \frac{45}{14} e^2+ 8 e^4 + \frac{685}{224} e^6 + \frac{255}{448} e^8 + \frac{25}{1792} e^{10}}{1 + 3 e^2 + \frac{3}{8} e^4}
\end{align}

For highly eccentric orbits, the dependence on the eccentricity reduces to
\begin{equation}
    \frac{(1-e^2)^{15/2}}{e^2(1-e)^9 g(e)} \to \frac{50}{(1-e)^{3/2}}.
\end{equation}
 The viscous time can be related to the lag time via 
 \begin{equation}
     t_\nu = \frac{3(1+2k_1)^2}{2k_1}\frac{R_1^3}{Gm_1\tau_L}\label{eq:t_visc}
 \end{equation}
thus, $\beta(e\to 1)$ is given by 
\begin{equation}
    \beta = \frac{3.17f_{\rm dyn} t_\nu}{ P_1  (1+2k_1)^2 (1-e)^{3/2}} 
\end{equation}

Setting $\beta=1$ and $k_1=0.014$ yields a transition eccentricity
\begin{equation}
    e_{\rm trans} = 1 -  \left(\frac{3f_{\rm dyn}t_\nu}{P_1} \right)^{2/3}, \label{eq:e_trans}
\end{equation}

Fig. \ref{fig2} shows $\beta(e)$ (Eq. \ref{eq:beta}) for various ratios of $P_1/t_{\nu}$.  The filled circles represent the approximate formula (Eq. \ref{eq:e_trans}). There is a good correspondence, therefore the qualitative behaviour can be undrestood using Eq. (\ref{eq:e_trans}). Generally, the transition eccentricity will approach unity once the inner period will increase. For a typical viscous time of $1\ \rm yr$ and wide binaries, $e_{\rm trans}$ is close to unity. For warm Jupiters and binaries around $\sim 5 \ \rm AU$, $e_{\rm trans}$ is around $0.8-0.9$\footnote{We note that for Jupiters at $\sim 1\ \rm au$, the viscous and orbital times are comparable, and dynamical tides could be dominant throughtout most of th eeccentricity range, and it required artificial truncation and better tidal models \citep{glanz_roz21, rozner2021}}. 

To summarize, taking $e_{\rm trans} \approx 0.8$ is consistent with planetary systems of warm Jupiters, where dynamical tides speed up the tidal dissipation beyond $e>e_{\rm trans}$, but for wider orbits the equilibrium tide model is the dominant one for much higher eccentricities compared with closer binaries.

\section{Numerical set-up and qualitative trends} \label{sec:3}
In order to explore the effect of wide triples coupling to galactic tide, we use \href{https://github.com/eugeneg88/SecuLab}{\texttt{SecuLab}}, a secular code which solves the equations of motion up to octupole order. In addition, the code includes additional effects of precession from GR and tidal bulges, and dissipation from both equilibrium and dynamical tides, and GW emission accorting to 2.5 Post-Newtonian expansion. We also include additional novel terms that correspond to single-averaging corrections for short-time variations of the outer orbit. The code is found in the public \href{https://github.com/eugeneg88/SecuLab}{GitHub repository}.

We first describe specific examples for CATGATE as to provide a detailed view on the type of secular and chaotic secular evolution through this process. We then describe our methods, and parameter space for a population synthesis study of CATGATE for low mass stars (below eight Solar masses, i.e. not explosing as core-collpase supernovae), including simplified stellar evolution considerations, tides, collisions and GR precession.

\subsection{Example of chaotic evolution} \label{sec:3.1}
The evolution of the inner binary is chaotic when the secular timescales are comparable. The ratio between the secular timescales is 
\begin{align} \label{r0}
    \mathcal{R}_0 & = \frac{\tau_{\rm sec}}{\tau_{\rm GT}} = \frac{2G\rho_{0}}{3\pi}\frac{m_{{\rm tot}}}{m_{{\rm out}}}\frac{P_{2}^{3}(1-e_{2}^{2})^{3/2}}{P_{1}} \\
    & = 1.16 \left( \frac{M_\odot}{m_{\rm out}} \right) \left( \frac{3 m_{\rm in} }{2 m_{\rm out}} \right)^{1/2} \left( \frac{200\rm au}{a_1} \right)^{3/2} \left( \frac{a_2}{2\cdot 10^4\rm au} \right)^{9/2} \biggr\rvert_{e_2=0}, \nonumber
\end{align}
where in the last row we assumed that $e_2=0$. Chaotic behavior is expected for $\mathcal{R}_0 \gtrsim 1$ \citep{hamers15,glp18}, or $a_1 \gtrsim 200\  \rm AU$. For these systems, the dominant conservative non Keplerian effect is GR precession from 1 PN terms. 

\begin{figure*} 
	\centering
	\includegraphics[width=8.5cm]{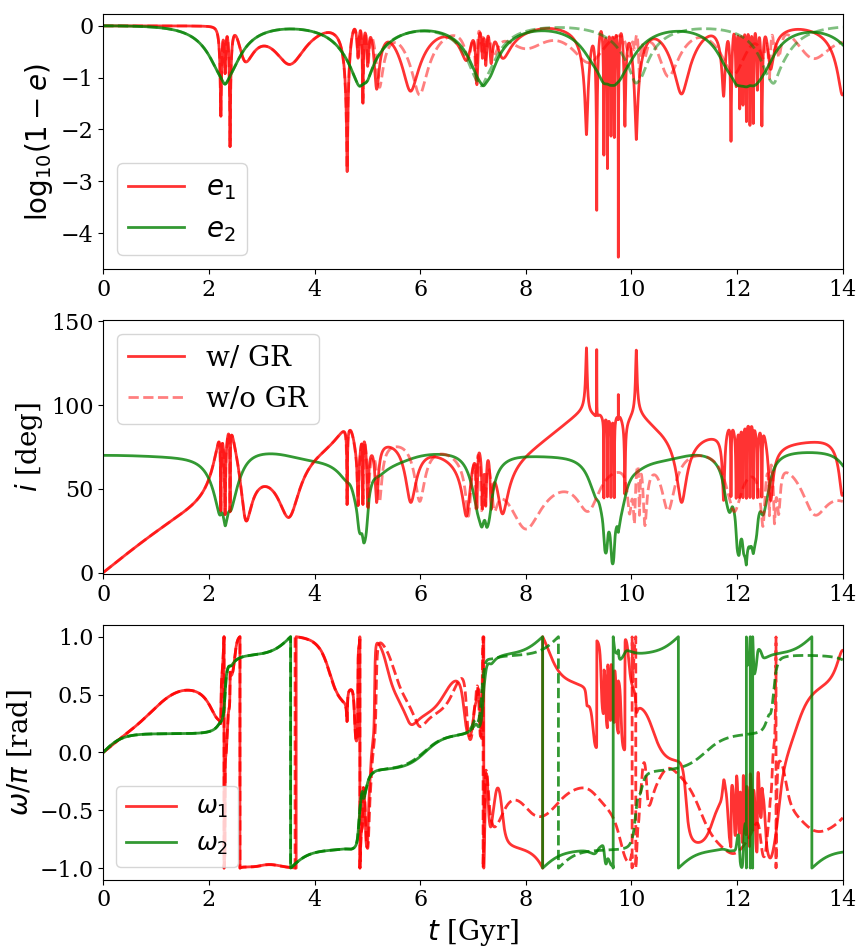} \includegraphics[width=8.9cm]{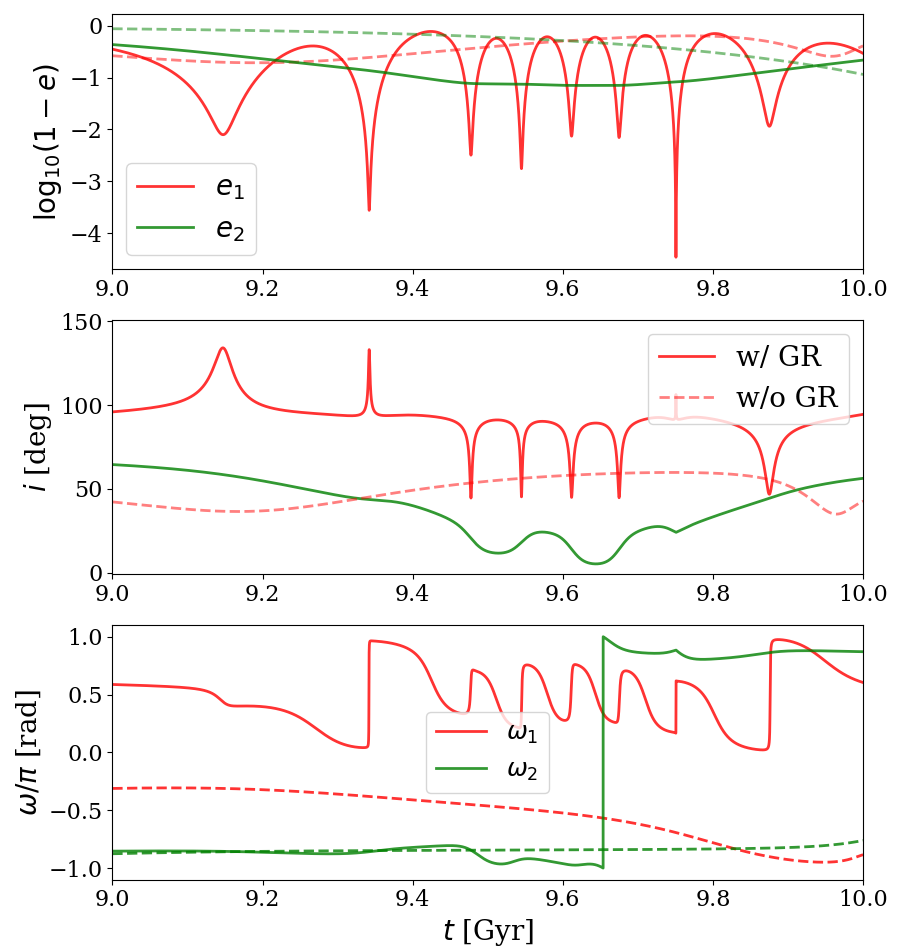}
		\caption{\label{fig1} Example of a chaotic CATGATE evolution. The initial conditions are $m_0 = m_1 = m_2 = M_\odot, e_1 = e_2 = 0.01, a_1 = 200\ \rm AU, a_2 = 2\cdot10^4\ \rm AU$. The initial inclinaion is $70^{\circ}$ to the Galactic plane. The apsidal and nodal angles are zero. The red lines represent the inner binary, while the green lines represent the outer binary. Solid lines represent the run with GR precession, while dashed lines represent purely Newtonian dynamics. Top: eccentricity. Middle: inclination between the planes of the inner binary A and the outer binary B. Bottom: Argument of pericentre. Left: Full evolution for $14$ Gyr. Right: The same evolution, zoomed in on the range of $9-10$ Gyr, where the inner eccentricity is excited to extreme values, the orbit experiences orbital flips and the argument of pericentre is resonant.}
	\label{fig:2}
\end{figure*}
Fig. \ref{fig:2} shows the evolution of a chaotic system, which consists of three equal Solar mass stars on almost circular orbits ($e=0.01$), separated by $200$ and $2\cdot 10^4\ \rm AU$, respectively, forming a hierarchical triple. The secular parameter $\mathcal{R}_0$ from Eq. \ref{r0} is of order unity, which indicates that strong chaotic excitations may occur. The initial triple systems is aligned, but it is misaligned with the Galactic plane by $70^{\circ}$.

We see that the outer binary oscillates every $2\ \rm  Gyr$ and drives the inner binary onto extremely eccentric orbit. Eventually at the forth cycle the eccentricity of the inner binary is extremely large, $1-e_1 \approx 10^{-4.5}$. We note that this large eccentricity is only present when we include extra precession terms, such as GR precession. 

The right panel of figure \ref{fig:2} shows the zoomed-in evolution between $9-10\ \rm Gyr$. We see that the argument of pericentre is caught in a transient librating state, which allows the rapid high eccentricity variations. 

We note that figure \ref{fig:2} is integrated with point sources without dissipative tides (only GW dissipation) or stopping conditions, and is merely an illustration of a proof of concept. Realistic stars with finite radii would have collided directly if the pericentre is low enough or interacted via mass transfer or CEE if the semi-major axis is reduced sufficiently via tidal interactions. We study a more realistic scenarios by means of population synthesis below.

\subsection{Population synthesis} \label{sec:3.2}
\subsubsection{Parameter space and initial conditions} \label{3.2.1}

\begin{table}
\begin{center}
\begin{tabular}{|c|c||c||c|}
\hline 
\multicolumn{4}{|c|}{{\bf Orbital Parameters}}\tabularnewline
\hline 
\hline 
$a_{1}$ & \multicolumn{3}{c|}{$\log\ U\sim[50,500]\ {\rm au}$}\tabularnewline
\hline 
$a_{2}$ & \multicolumn{3}{c|}{$\log\ U\sim[5\times10^{3},5\times10^{4}]\ {\rm au}$}\tabularnewline
\hline 
$e_{1},e_{2}$ & \multicolumn{3}{c|}{thermal: $ f(e)\propto2e$}\tabularnewline
\hline 
$i_{1},i_{2}$ & \multicolumn{3}{c|}{cosine Uniform}\tabularnewline
\hline 
$\Omega_{i},\omega_{i}$ & \multicolumn{3}{c|}{Uniform in $[0,2\pi)$}\tabularnewline
\hline 
\multicolumn{4}{|c|}{{\bf Physical properties}}\tabularnewline
\hline 
$m_{1},m_{2}$ & \multicolumn{3}{c|}{$\begin{matrix}{\rm \text{Low Mass: Kroupa (0.5,1)}}\\
{\rm \text{Int. Mass: Kroupa (1,8)}}
\end{matrix}$}\tabularnewline
\hline 
$m_{3}$ & \multicolumn{3}{c|}{Kroupa (0.5,1)}\tabularnewline
\hline 
$R_{i}$ & \multicolumn{3}{c|}{$\begin{matrix}(m_{i}/M_{\odot})^{\xi}R_{\odot}\ {\rm MS}\\
10\times R_{i}(0)\ {\rm RG}\\
0.01R_{\odot}\ {\rm WD}
\end{matrix}$ }\tabularnewline
\hline 
$k_{1}$ & \multicolumn{3}{c|}{$\begin{matrix}0.014\ {\rm MS}\\
0.1\ {\rm Other}
\end{matrix}$}\tabularnewline
\hline 
$t_{\nu}$ & \multicolumn{3}{c|}{$\begin{matrix}5\ {\rm yr}\ {\rm MS}; 10^4\ {\rm yr\  if} M>1.25M_\odot \\
0.7\ {\rm yr}\ {\rm RG}\\
10^{7}\ {\rm yr}\ {\rm WD}
\end{matrix}$}\tabularnewline
\hline 
$f_{\rm dyn}$ & \multicolumn{3}{c|}{0.03}\tabularnewline
\hline
\end{tabular}
\[
\]
\par\end{center}
\caption{Initial distributions for the orbital and physical parameters.}\label{tab:ic}
\end{table}

Since the binaries are generally wide, their orbital parameters are assumed to be uncorrelated. The masses are drawn from the low mass Kroupa mass function, $dN/dm\propto m^{-2.3}$. Once the mass is determined, the radius is determined from the mass-radius relation $R/R_\odot = (m/M_\odot)^\xi$ with $\xi=0.8$ for $m<M_\odot$, and $\xi=0.57$ for $m>M_\odot$ \citep[e.g.][]{MRR}.  The eccentrities are drawn from a thermal distribution, $dN/de\propto 2e$. The separations are drawn from log-uniform distribution, $dN/da \propto 1/a$. The inclinations are cosine-uniform $dN/d\cos i \propto 1$. The nodal and apsidal angles are drawn uniformly, $\omega,\Omega \sim U[0,2\pi]$. The latter are generally motivated by observations \citep{moe-distefano2017}.

For MS stars, we use the apsidal motion constant $k_1=0.014$ and a viscous time of $t_{\nu} = 5\rm \ yr$. Stars above $m\gtrsim1.25M_\odot$ have radiative envelopes, and their radiative tides are much weaker.

Generally, most population synthesis studies rely on detailed prescriptions which rely on stellar evolution codes to determine $k_1$ and the ratio $k_1/T_c$ where $T_c$ is related to the tidal dissipation \citep{hurley2002, hamers21}. In our case, tides are important only when the pericentre is close, while we aim to estimate the fraction of systems that can reach low pericentre to begin with. Hence, the role of tides is secondary and the typical parameters are taken for an order of magnitude.

\subsubsection{Simplified stellar evolution} \label{sec:3.3}

The main-sequence lifetime of stars is empirically approximated as (e.g. \citealp{hansen2004})
\begin{equation}
    t_{\rm ms} = 10 \left(\frac{m}{M_\odot}\right)^{-2.5}\rm Gyr \label{t_ms}
\end{equation}
which is comparable to our secular and chaotic evolution timescales for star above a Solar mass. The red giant phase has a timescale of $10-15\%$ of the main sequence lifetime. We set the time for WD formation as $t_{\rm WD} = 1.15t_{\rm ms}$ and the radius to be $10$ times the initial radius. This is in no way a detailed evolution model but just a proxy to probe the importance of this channel to red giant binary interactions.

The apsidal motion constant of red giant is increased, and we set it to $k_1=0.1$ from inspecting the tables of \cite{claret1992}. The viscous time is set for simplicity to be $t_{\rm \nu}= 5 (k_1/0.014)^{-1} \ \rm yr$ to maintain a roughly constant lag time, and is accurate to within an order of magnitude within more detailed prescriptions \citep{hurley2002, hamers21}.

The WD phase is the final stage of evolution for the low-mass stars we consider. The WD radius is is taken to be $10^{-2}R_\odot$. The apsidal constant is $k_1=0.1$ estimated from \cite{vila1977}, and the viscous timescale is large, $t_{\rm \nu} \gtrsim 10^7\ \rm yr$ \citep{campbell84}. We expect equilibrium tides to be negligible in the WD phase due to the small radii and large viscous times, but dynamical tides could still play a role in highly eccentric encounters.

In any case, the strength of the dynamical tide is $f_{\rm dyn}=0.03$, which is more appropriate for MS stars \citep{mk18}. Table \ref{tab:ic} summarizes the initial conditiosn and distributions of the parameters.

\subsection{Stopping conditions}\label{3.3}
The stopping conditions are: 

i) The time reaches $t_{\rm end}=10\ \rm Gyr$. 

ii) The pericentre is a few times the mutual radius $r_p=\zeta (R_1+R_2)$. 
We choose $\zeta=3$, which is close enough to the value of $2.7$ from detailed tidal disruption simulations \citep{g13}.

iii) The semi-major axis shrinks to $10\%$ of its initial value. 

iv) The triple becomes dynamically unstable according to the \cite[][MA]{ma01} stability criterion

\begin{equation}
\frac{a_2(1-e_2)}{a_1} \le 2.8 \left[ \left(1 + \frac{m_{\rm out}}{m_{\rm in}} \right) \frac{1+e_2}{(1-e_2)^{1/2}} \right]^{2/5}
\end{equation}

These stopping consitions correspond to four possibilities: i) No significant interaction occurs during the evolution. ii) Close encounter, probably a tidal disruption or direct collision. iii) Efficient tidal migration.
 iv) Dynamical instability according to the MA criterion.

\begin{figure*}
    \centering
    \includegraphics[width=7cm] {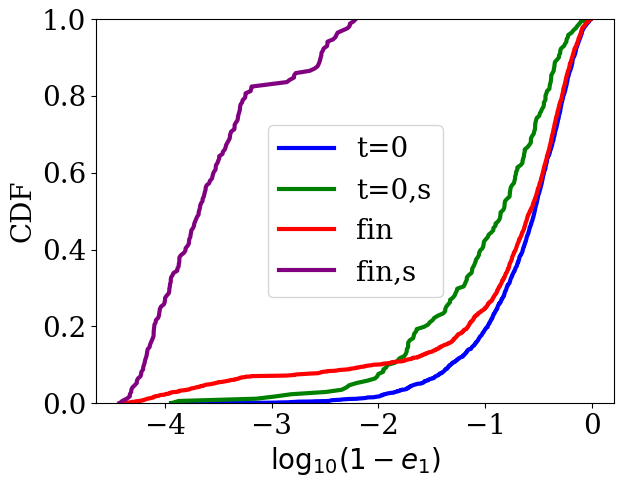}\includegraphics[width=7cm] {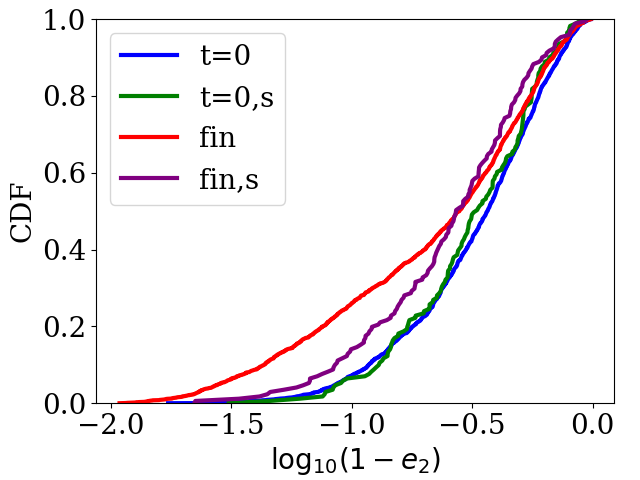}
    \caption{\textbf{Distributions of initial
and final eccentricities}. The blue lines indicate the initial eccentricity distribution of all runs ($t=0$ label). The green lines show the distributions of the runs that stopped once the critical separation for the pericentre has been reached ($t=0,\rm s$ label). The red lines show the final eccentricity distribution for all runs ($\rm fin$ label).The purple lines show the final eccentricity distribution of the runs that stopped once the critical separation for the pericentre has been reached ($\rm fin, s$). The left panel shows the inner eccentricity $e_1$, while the right panel shows the outer eccentricity. CDF is given in log scale of $1-e$.}
    \label{fig:es}
\end{figure*}

\begin{figure*}
    \centering
    \includegraphics[width=7cm] {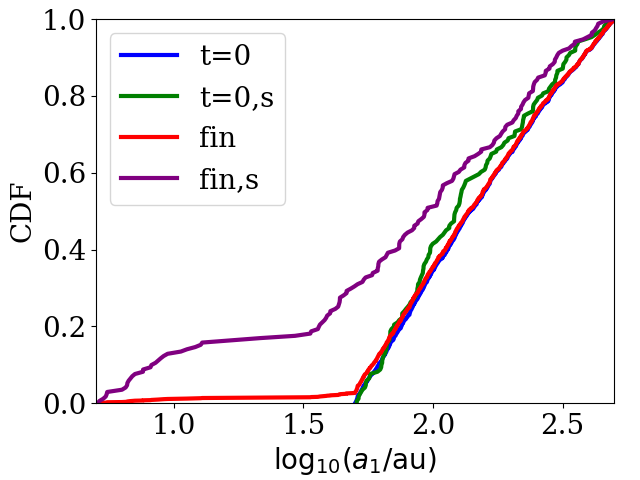}\includegraphics[width=7cm] {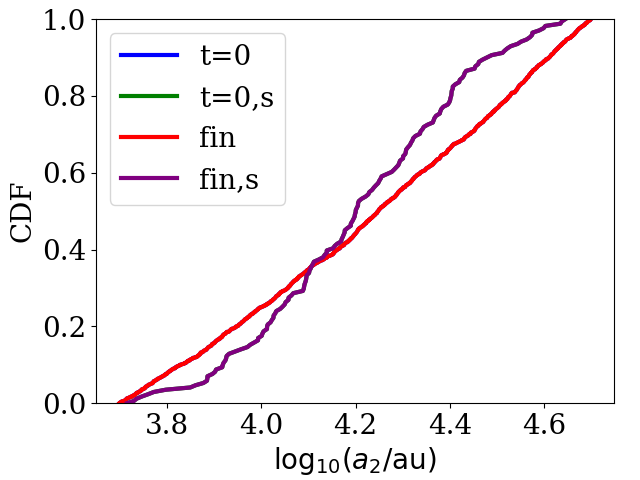}
    \caption{Same as Fig. \ref{fig:es}, but for $a_1$ and $a_2$.}
    \label{fig:as}
\end{figure*}

\section{Population synthesis results} \label{sec:4}

\subsection{Main sequence stars} \label{sec:4.1}

In order to explore the fraction of wide triple systems that undergo chaotic evolution we initialize 2000 systems as follows: The initial conditions are drawn as described in sec. \ref{3.2.1}. The boundaries are $=50\ \rm au < a_1 < 500\ \rm au$, and $5000\ \rm au < a_1 < 50000\ \rm au$. The masses follow the Kroupa mass function between $0.5\ M_\odot <m< 1 M_\odot$.  We run each system up to a time of $t_{\rm end} = 10\ \rm Gyr$ or until one of the stopping conditions applies, as described below. The output recorded every $10^4\ \rm yr$, for a total of $10^5$ data points per simulation.

\subsubsection{Statistical properties}

Out of the $2000$ simulations,  $\sim27\%$ ($538$) are unstable, and are expected to lead to an ejection of one of the stars\footnote{A fraction of $\sim30\times R/a_{\rm in}$ could still result in a close inteactions, where R is the stellar radii at the time and $a_{\rm in}$ is the inner binary sma ; see e.g. discussion of collisions in an unstable system in \cite{pk12}. Given the small ratio, we neglect such interactions in the curent study, which at most get to a few percents for RGs.}. Around $7.2\%$ ($144$) end up with pericentre smaller than our threshold. Since they are extremely eccentric, they enter the Roche limit on almost a radial orbit and probably will be tidally disrupted or collide. Around $1.3\%$ ($26$) and up with a separation ten times smaller the initial value, due to efficient tidal dissipation. These systems will probably be tidally circularized and eventually form close binaries. The rest of the sample remained stable and no significant interaction occurred. 

\begin{figure*}
    \centering
    \includegraphics[width=7cm] {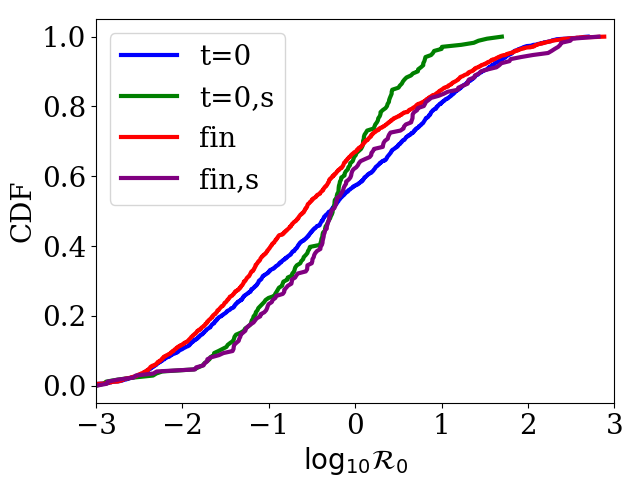}\includegraphics[width=7cm]{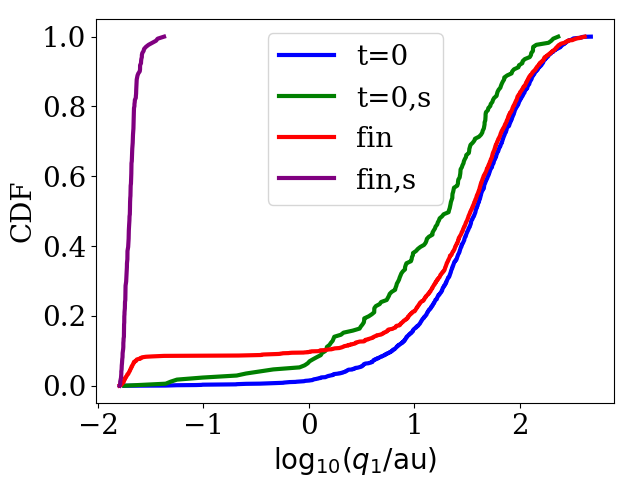}
    \includegraphics[width=7cm] {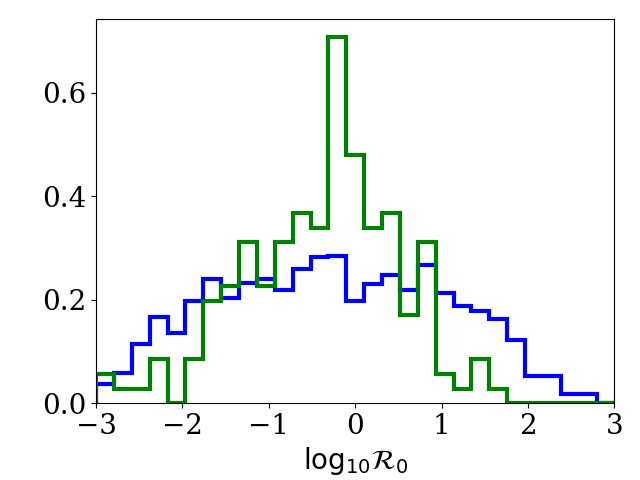}\includegraphics[width=7cm]{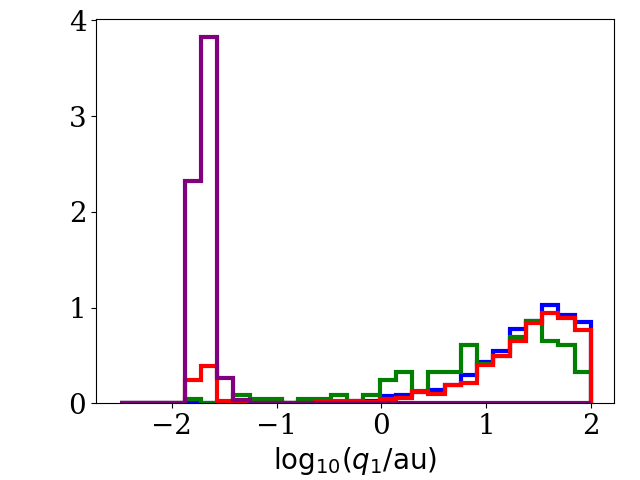}
   
    \caption{Same as Fig. \ref{fig:es}, but for the parameter $\mathcal{R}_0$ (left) and the pericentre $q_1$ (right). The solid lines indicate the recorded pericentre at the end of the run, while the dashed lines correpond to the minimal recorded pericentre. The bottom panels show the density plot of the data of the top panels. }
    \label{fig:R-q}
\end{figure*}

The fraction of unstable systems is derived in Appendix A and is equal to $f_{\rm U}(e_f) = 2\sqrt{(1-e_f^2)/5}$ where $e_f$ is the eccentricity above the system is typically unstable. Since the mean separation ratio is $\langle a_2/a_1\rangle  \approx 100$ and instability occurs roughly when $a_2 \approx 5a_1$, a good choice for $e_f$ is $e_f=0.95$, which gives $f_{\rm U}\approx 28\%$, or $\sim 557$ unstable systems, which is remarkably close to the recorded value of $538$.

In what follows, we show the initial and final cumulative distribution function (CDF) of various parameters. We distinguish between the entire sample 'all', and the sample of systems that were stopped due to conditions ii) and iii) 'terminated', or 'stopped' samples. Stopping due to dynamical instability is not considered in the stopped sample because it will most likely lead to an ejection and not to close encounter.

Fig. \ref{fig:es} shows the CDF of the  eccentricities. The left panel shows the CDF of the inner eccentricity $e_1$ (or the CDF of $\log_{10}(1-e_1)$, so large negative values indicate values closer to unity, $e_1\to1$.). The initial inner eccentricity $e_1$ follows a thermal dsitribution (blue line). The sample of the initial eccentriciy of orbits which were terminated (green line) is slighly above thermal, but not significantly.  

The final inner eccentricity CDF (red line) is more eccentric. The vast majority of the extreme eccentricies ($\log_{10}(1-e_1) < -3$) is achieved predominantly in terminated orbits, and a large eccentricity  ($\log_{10}(1-e_1) < -2$) is a requirement for close encounter, as expected. The eccentricities are also pushed to the extreme values of $1-e_1 \approx 10^{-4.5}$, which are unachievable by initial conditions for standard ZLK oscillations.

\begin{figure*}
    \centering
    \includegraphics[width=5.9cm]{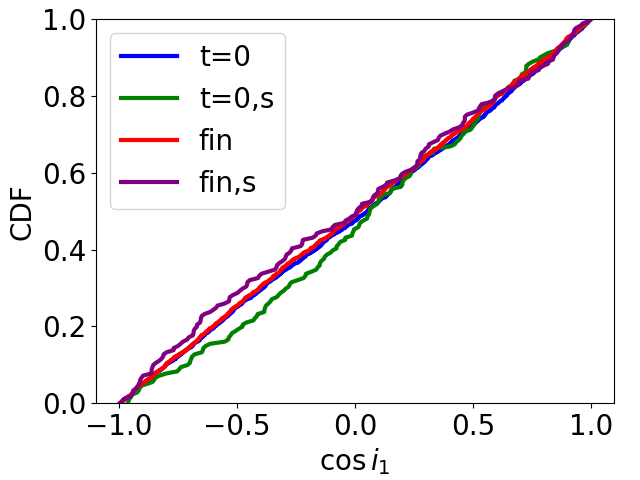}\includegraphics[width=5.9cm]{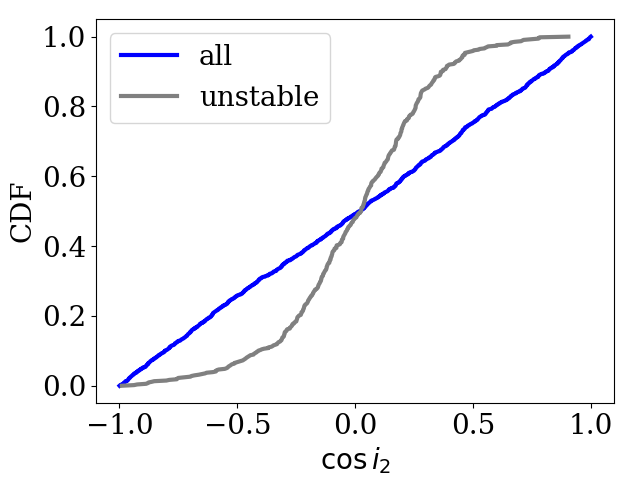}\includegraphics[width=5.9cm]{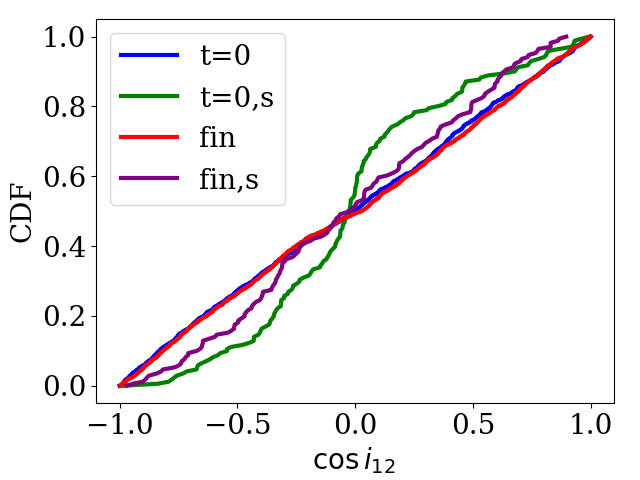}
      \includegraphics[width=5.9cm]{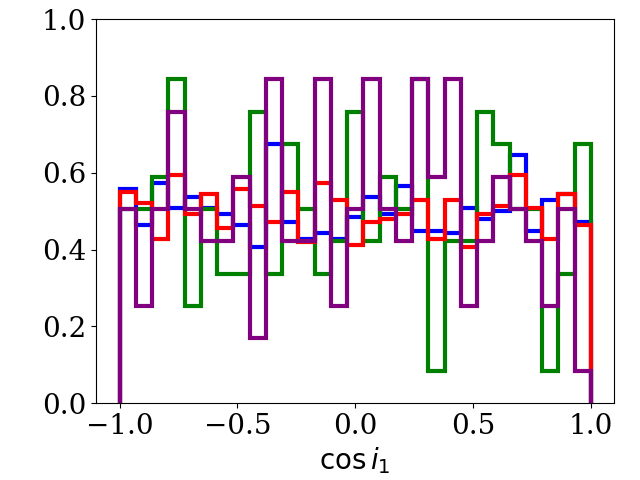}\includegraphics[width=5.9cm]{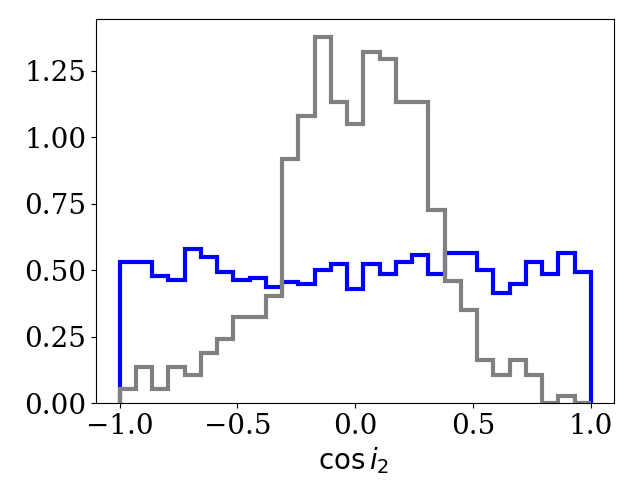}\includegraphics[width=5.9cm]{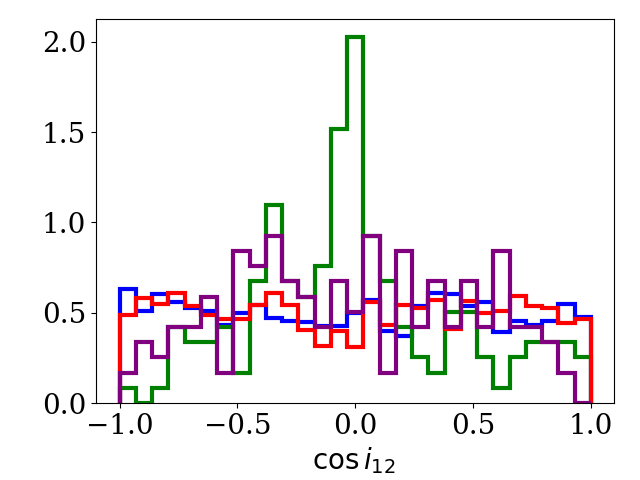}
    
    \caption{Same as Fig. \ref{fig:es}, but for the inner (left), outer (middle) and mutual (left) inclinations.  The middle panel shows the initial outer inclination that eventually rendered unstable (grey), together with the entire initial sample (blue). }
    \label{fig:incs}
\end{figure*}

The right panel shows the CDF of the outer eccentricity. Here we also see a trend that the initial outer eccentricity $e_2$ that leads to termination is slightly more eccentric than the initial distribution. The initial and final 'all' $e_2$ are truncated at around $e_2\lesssim 0.98$ to allow dynamical stability. The final $e_2$ is on average more eccentric, but the maximal value  for terminated orbits is somewhat lower, around $e_2\approx 0.95$, to allow dynamical stability. We'll show later in Fig. \ref{fig:incs} that dynamical instability is predominantly achieved when the outer inclination angle $i_2$ is initially close to $\pi/2$.   

Fig. \ref{fig:as} shows the CDF of the separation. The left panel shows the inner separation. The initial separation is distributed as log-uniform (blue), as do the distribution of terminated orbits. The final distribution (red) is also almost log-uniform, up to a tail of small separations for orbits that undergone efficient dissipation. The final distribution of terminated orbits is more skewed towards lower separations; $\sim 40\%$ of the terminated orbits end up with $a_1<50\ \rm au$, while the orbits that have migrated to smaller separations include roughly a quarter of the terminated orbits. This means that some orbits tidally evolved before reaching small pericentre and terminating.

The right panel of Fig. \ref{fig:as} shows the outer separations. The initial and final separations are essentially identical both for all the initial conditions and the terminated ones. The orbits that lead to termination are slightly more skewed, partially due to the  absence of chaotic evolution for the very small or very large separations.

In order to explore the role of the chaotic dynamics we plot the CDF of the $\mathcal{R}_0$ parameter (Eq. \ref{r0}) in the top left panel of Fig. \ref{fig:R-q}, while the density plot (a histogram where the area is normalised to 1) is shown in the bottom left panel. The initial distribution is roughly uniform in the range $-3 \le \log_{10}\mathcal{R}_0 \le 3$ (blue lines), however the initial conditions that lead to termination are predominantly between $-0.3 \le \log_{10}\mathcal{R}_0 \le 0.7$; $\sim50\%$ of the terminated systems are in this range. The final recorded $\mathcal{R}_0$ also follows roughly uniform distribution, although in the terminated orbits there is a dearth of final orbits around $\log_{10}\mathcal{R}_0 \ge 0$. The reason is probably due to reduction of the outer eccentricity $e_2$ at the close encounter, which reduces $\mathcal{R}_0$. The systems which are more prone to strong encounters are the ones in the chaotic regime where $\mathcal{R}_0 \sim \mathcal{O}(1)$ and with large misalignment in respect to  the Galactic plane, which can give rise to large variations in $e_2$.
The bottom left panel shows only the initial $\mathcal{R}_0$, showing that the terminated orbits prediminantly originate at $0.5\lesssim \mathcal{R}_0 \lesssim5$

The top right panel of Fig. \ref{fig:R-q} shows the CDF of the pericentre $q_1= a_1(1-e_1)$, while the bottom left panel shows the density plot.  While there is little dependence on the initial pericentre between the total CDF and the terminated CDF, the vast majority of the terminated systems have a pericentre smaller than a few Solar radii. The maximal pericentre of a terminated orbit is $\approx 0.04\rm \ au$, which means that low pericentre is required for both radial plunge and efficient tidal dissipation. The bottom right panel shows that a low $q$ is required for any significant interaction.

The top rows of figure \ref{fig:incs} shows the CDF of the initial and final inclinations, while the bottom rows show the density plots of the same data. The left panels show the inner inclination $i_1$ with respect to the Galactic plane. We see little variation in $i_1$, which means that the evolution is mostly independent on $i_1$. The middle panels show the inclination of the outer binary with respect to the Galactic tide $i_2$. The total initial and final $i_2$ doesn't change much, but the initial $i_2$ which leads to instabilities (grey line) is located mostly around $\pi/2$. This is to be expected since $i_2$ close to $\pi/2$ leads to high excitation of $e_2$ which destabilizes the triple system. The right panels show the mutual inclination $i_{12}$. We see that the initial and final values don't change much, but the initial value of terminated orbits is more clustered around $i_{12}=\pi/2$. These orbits could interact on shorter timescale without the aid of the Galactic tide, as we demonstrate in the next section.

\subsubsection{Correlations}
\begin{figure*}
    \centering
    \includegraphics[width=8.1cm]{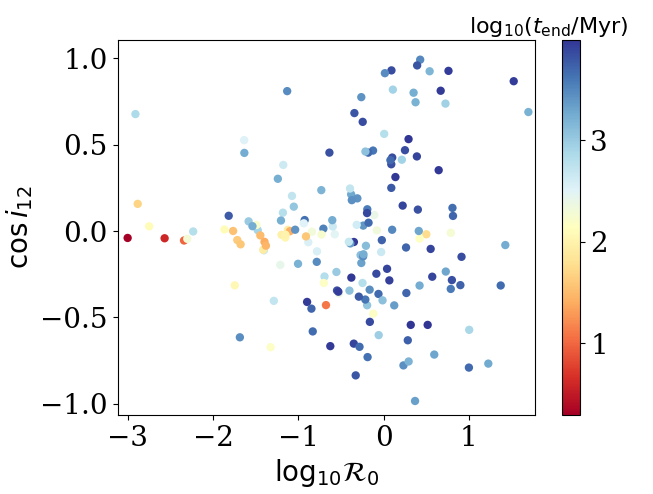}\includegraphics[width=8.5cm]{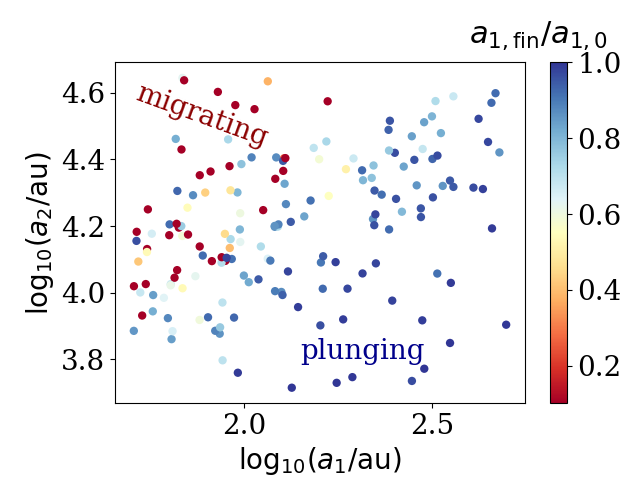}
    \caption{Scatter plots of various parameters. Left: cosine of the mutual inclination $i_{12}$ versus the (logarithm of the) $\mathcal{R}_0$ parameter, with the stopping time as a color code. Right: log-log plot of the initial outer versus initial inner separations, with the final separation in units of the initial separation as a color code.}
    \label{fig:low_mass_scatters}
\end{figure*}

In order to explore correlations between various parameters, we show in Figure \ref{fig:low_mass_scatters} the scatter of key parameters of the terminated orbits. The left panel shows the scatter of the mutual inclination and the $\mathcal{R}_0$ parameter. Red marks indicate short stopping times while blue marks indicate longer stopping times. We see that the orbits that were terminated first are close to being initially polar, i.e. $\cos i_{12} \approx 0$. Their low value of $\mathcal{R}_0$ suggest that they have a relative short secular timescale, on the order of $10-100\ \rm Myr$, comparable to their stopping time. These triples evolved via isolated triple evolution without the aid of the Galactic tide. Only at $\mathcal{R}_0\approx 0.1$ the scatter in $\cos i_{12}$ starts to widen and the stopping times increase to be $t>1\ \rm Gyr$, suggesting the chaotic Galactic tidal evolution took place.

The right panel of Figure \ref{fig:low_mass_scatters} shows the log-log scatter in the initial inner and outer separations, where the color code is the relative shrinking of the inner orbit. There are two distinct populations, the migrating one (which stopped at $a_{1,\rm fin} = 0.1 a_{1,0}$) and the plunging one, which is essentially partially migrating, or not migrating at all, but where the stopping was due to small pericentre approach. The two populations are distinguished mainly by their initial separation; the ones with $a_1\lesssim 100\ \rm au$ is affected by additional precession terms from GR and tides that limit the maximal eccentricity and save it from disruption and allow efficient migration, while systems with orbits with $a_1 \gtrsim 100\ \rm au$ have larger unconstrained eccentricities and tend to plunge and probably be disrupted or end in direct collision, pending on the specific types of stars nad mass-ratios.

\begin{figure*}
    \centering
    \includegraphics[width=5.8cm]{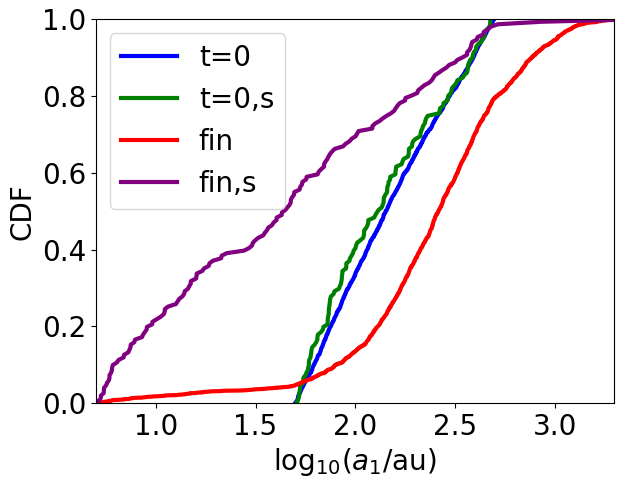}\includegraphics[width=5.8cm]{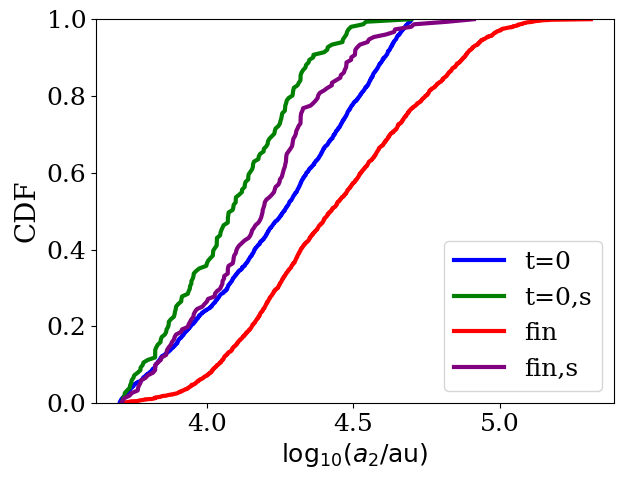}\includegraphics[width=5.8cm]{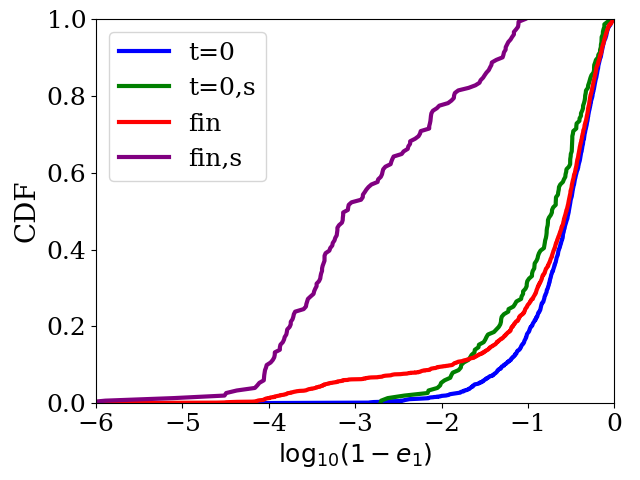}
    
    \caption{Same as Fig. \ref{fig:es}, but for the RG runs. Left: inner separation. Right: outer separation. Right: inner eccentricity. }
    \label{fig:9}
\end{figure*}

{\bf To summarize}, $27\%$ of the triples are dynamically unstable and most likely become unbound. $7.2\%$ become close binaries with their pericentre is less than 3 times their mutual stellar radii, while another $1.3\%$ are efficiently shrinking due to efficient orbital energy loss. 
Most of the close encounters are induced by lowering the pericentre of the outer binary due to the Galactic tide and triggering chaotic evolution.

\subsection{Red Giants and Common envelope evolution} \label{sec:4.2}

In order to add more massive stars that undergo stellar evolution within the duration of the dynamical simulation we must specify the details of stellar evolution and mass loss prescription. We assume an adiabatic mass loss, namely that $m/\dot{m} \gg P_2$, which is roughly correct for the widest binaries, providing that stars are not too massive. In this case the angular momentum is conserved and we have for a star with initial mass $m_0$:
\begin{equation}
    m(t) = m_0 \left( \frac{m_0}{m_{\rm WD}} \right) ^ {-\frac{t-t_{\rm MS}}{t_{\rm WD}-t_{\rm MS} } };\quad t\in (t_{\rm MS}, t_{\rm WD}) \label{eq:m_of_t}
\end{equation}

Note that the vast majority of our inner binaries are initially sufficiently wide such we can assume no significant mass-transfer or interaction occurs between the inner binary components, in the absence of CATGATE effects.

For the final WD mass we use the empirical relationship of \cite{catalan2008}:
\begin{equation}
    m_{\rm WD}(m)=\begin{cases}
 0.096 m + 0.429 & m<2.7\\
0.127 m + 0.31 & m>2.7
\end{cases}\label{eq:mwd}
\end{equation}
where $m,m_{\rm WD}$ are given in units of $M_\odot$. 

Although it is not obvious at first glance, the mass is exponentially decaying $m(t) \propto \exp\left( (t-t_{\rm MS})/\tau_{\rm ML} \right)$ with the mass loss time-scale being $\tau_{\rm ML} = (t_{\rm WD} - t_{\rm MS})/\ln (m_0/m_{\rm WD})$. In order to conserve the angular momentum, the orbits expand as 
\begin{align}
    a_1(t) = & a_{1,\rm in} \frac{m_{1,\rm in} + m_{2, \rm in}} {m_1(t) + m_2(t)} \\
    a_2(t) = &a_{2,\rm in} \frac{m_{1,\rm in} + m_{2, \rm in} + m_{3,\rm in}} {m_1(t) + m_2(t) + m_3(t)} \label{eq:expansions}
\end{align}
where it is assumed that at $t>t_{\rm WD}$ the mass is $m_{\rm WD}$ for each star. The MS timescale $t_{\rm MS}$ is given in Eq. (\ref{t_ms}). For the WD time scale $t_{\rm WD}$ we use the rule of thumb that the RG phase is $\sim 10-15\%$ of the MS timescale, so we set $t_{\rm WD}=1.15t_{\rm MS}$. At the RG branch, we keep the radius to be conservatively $10$ times the initial value. The tidal parameters are described in the initial conditions section. In principle, such mass-loss evolution could lead to instbaility of the triple system (triple evolution instability; \citealp{pk12,ham+21,too+21}) in a small fraction of the cases, and can affect the type of secular and quasi-secular evolution \cite[e.g][]{pk12,sha+13,michaely2014, toonen2018}.

\subsubsection{Combined effects of stellar evolution and galactic tides} 

\begin{figure*}
    \centering
    \includegraphics[width=8.1cm] {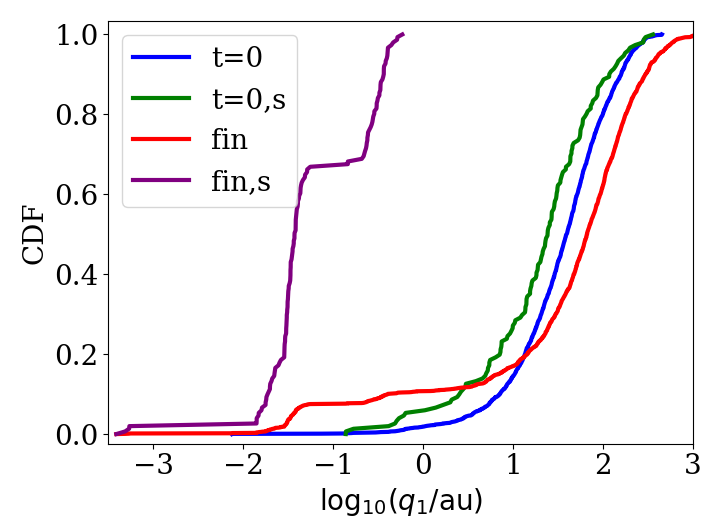}
    \includegraphics[width=8.0cm]{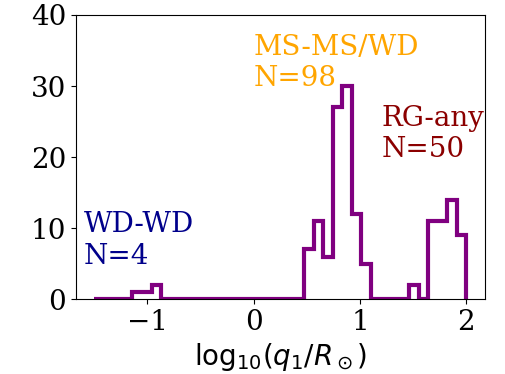}
    
    \caption{Left: Same as Fig. \ref{fig:R-q} but for the RG run and only for the pericentre $q_1$ parameter $\mathcal{R}_0$ Right: The histogram of the final pericentre of the stopped orbits. Note that this is a histogram, not a density plot and $q_1$ in normalised to Solar radii, not au. Each cluster is indicated with the potential progenitors of the close encounter. }
    \label{fig:10}
\end{figure*}

\begin{figure*}
    \centering
    \includegraphics[width=9.5cm] {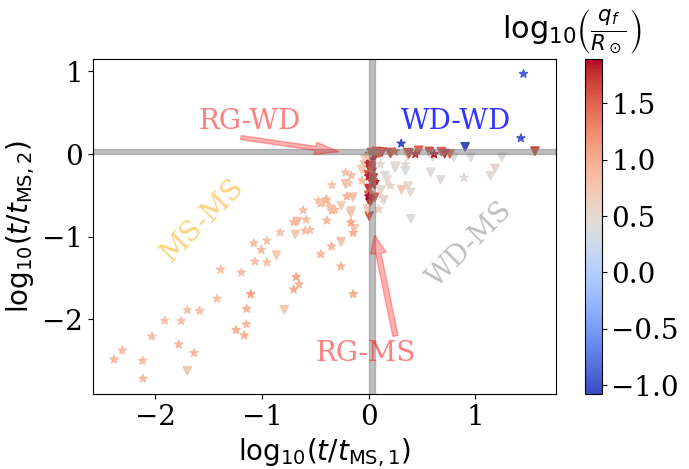}\includegraphics[width=7.8cm]{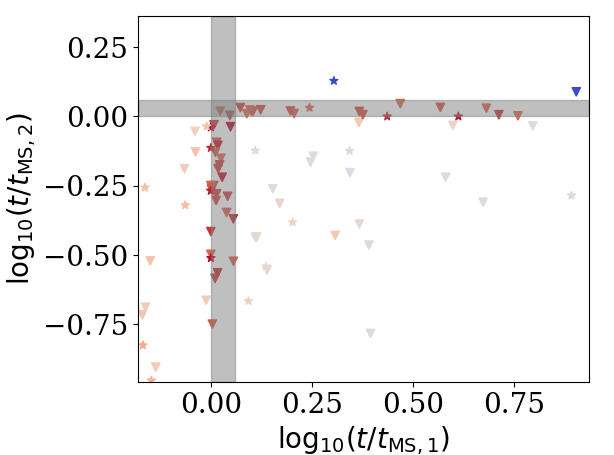}
    
    \caption{Typical timescales for the merger in terms to the stars' MS lifetime. The X and Y axes are stopping time in units of $t_{\rm MS,1}$ and $t_{\rm MS,1}$, respectively. Since we set up $m_1>m_2$, $t_{\rm MS,1}$ is always shorter than $t_{\rm MS,2}$. The grey areas indicate the time where $m_1$ is a RG (vertical area) and where $m_2$ is a RG (horizontal area). Bottom left square corresponds to the bulk of MS-MS collisions, bottom right area corresponds to MS-WD collisions. Top right corner corresponds for WD-WD collisions. The color code is the pericentre at the soptting time. The blue dots terminate with low pericentre, indicative to WD collisions. Stars indicate systems with low q (disruptions or mergers. low $q_1$), while triangles indicate migrating systems (low $a_1$. The dark red dots indicate large pericentre indicative for RG encounters. Right panel shows the zoomed in region near the RG evolution time.  }
    \label{fig:11}
\end{figure*}

In order to explore the role of stellar evolution and the contribution of red giants to close encounters, we ran 2000 additional simulations where the two inner stars are drawn from the same mass function $dN/dm \propto m^{-2.3}$, but in the range $1 \le m/M_\odot \le 8$. We also assign the more massive star to be $m_1$, such that it evolves faster, i.e. $t_{\rm MS}(m_1) < t_{\rm MS}(m_2)$. Out of the 2000 simulations around $41.5\%$ (831) became dynamically unstable. Another $3.6\%$ (72) ended up with low pericentre and $4\%$ (80) ended up with low separation. 

\begin{table}
\begin{center}
\begin{tabular}{|c|c|c|c|} 
 \hline
  & MS & RG & WD \\ [0.5ex] 
 \hline\hline\
 MS & 71 (47\%) & 29 (19\%) & 27 (18\%) \\ 
 \hline
 RG &  & 2 (1.3\%) & 19 (12.5\%) \\
 \hline
 WD &  &  & 4 (2.7\%) \\ [1ex] 
 \hline
\end{tabular}
\end{center}
\caption{Total number of close encounters for different stellar types. The percentage is of the total number of encounters. }\label{tab:1}
\end{table}

Figure \ref{fig:9} shows the distributions of the separations and the inner eccentricity. We note that the inner separation (left panel) undergoes both expansion due to mass loss in the RG phase and contraction due to close encounters and tidal evolution. The outer separation (middle panel) undergoes expansion. For the inner eccentricity, the distribution is similar to the lower mass stars case, but now there are two more clusters for the final treminated orbits (purple line): eccentricities between $0.1 \gtrsim 1-e\gtrsim 0.01$ for RG close encounter, and extreme eccentricities below $\log_{10}(1-e)=10^{-5}$ for direct WD-WD collisions.

Figure \ref{fig:10} shows the distribution of the pericentre. The top panel shows that the final pericentre of the stopped systems $q_1$ has a trimodal distribution. This is evident in the histogram in the bottom panel. Note that $q_1$ in normalised to units of solar radius. The bulk of the systems merge before stellar evolution kicks in, or later when star 1 is a WD and star 2 is still on the MS (98), which gives a pericentre of a few Solar radii. Around a third of the stopped orbits (50) stop when one of the stars is a RG, leading to several dozends of Solar radii separation. A handful of systems (4) were able to collide as binary WD, when the pericentre is below a tenth of a Solar radii. 

Table \ref{tab:1} further lists the number of terminated systems (and percentage) for all of our stellar types. We see MS-MS encounters are the most common, with nearly half the systems ending in this state. RGs most likely encounter MS stars in about $1/5$ of the systems and with WDs in about $1/8$ of the systems, and only a handful are RG-RG encounters. This is mainly due to the nearly equal mass of the binary components progenitors required in order for both to be on the RG at the time of the close encounter. WD-WD collisions are relatively rare compared with WD-RG or WD-MS close encounters.

In order to explore the effects of stellar evolution we plot in Figure \ref{fig:11} the stopping times normalized to the MS timescales of stars 1 and 2, respectively. Mass $m_1$ is always larger than $m_2$, thus $t_{\rm MS,1}$ is smaller than $t_{\rm MS,2}$, thus the upper left triangle is forbidden. We see that in the bottom left corner, the MS-MS encounters occur when the pericentre is a few Solar radii, around $10$ or slightly less. Once the first stars evolve to become RGs, there is a pile-up on systems that merge in the RG phase within the vertical gray area (it is better visualized in the zoomed in right panel). The other star may also evolve to RG which would lead to RG-RG encounters (small grey box in the zoomed in plot), but it is less likely and only two system experience it. In most cases, the first star evolves into a WD and then encounters the MS conpanion, which results in a pericentre around a factor of $\sim 2$ lower than for MS-MS collisions or around $\sim 5$ solar radii. Some systems will have an encounter when the second star evolves into a RG which leads to RG-WD encounters (horizontal grey line). These could then lead to CEE. Finally, a small fraction of systems stop when both stars are WDs and have very small pericentre, which will most likely lead to direct WD collisions and possibly Type Ia SNe.

The systems that end with low $q_1$ (disruption or collision) are most likely to be MS-MS or WD-WD collisions. This is also evident from sec. \ref{sec:4.1}, where the vast majority of the terminated systems were with low $q_1$. This occurs since for wide binaries the role of tidal dissipation in MS and in WD stars is relatively weak. For the cases of RG close encounters, the bulk of the terminated systems had efficiently migrated and terminated with low $a_1$. This is most likely due to stronger tidal dissipation owing to the large radii and convective envelopes of the RGs.

{\bf To summarize}, around $3.6\%$ of systems end up with low-pericentre and another $4\%$ efficiently migrate inwards. The efficient migration occurs due to the large radius of the convective RGs, while MS and WD encounters are generally concluded as tidal-disruptions or collisions. Around two thirds (or $4.9\%$ of the total systems) are MS-MS or MS-WD collisions. Another third (or $2.5\%$ of the total systems) are RG collisions with any stellar type, and a handful (or $0.2\%$) of the systems are WD-WD collisions.

\section{Discussion} \label{sec:5}

\subsection{Rates of close encounters} \label{5.1}
 
 Here we provide an approximate (given the simplifications made) estimate for the rate of strong interactions between the inner binary components catalyzed by the CATGATE process. 

Assuming the number of stars in the Galaxy is $N_\star$. A fraction $f_b$ of them are in binaries with log uniform separation spaning five orders of magnitude between $0.5\ \rm au$ and $50,000\rm au$. For a wide triple the inner binary should be within the $(50,500)\ \rm au$ octave, and the outer should be within the $(5,50)\times 10^3\ \rm au$ octave. The hierarchical structure and the ultra-wide outer binary companion are unlikely to affect the binarity of the inner orbit. If the distributions are uncorrelated,  we have $N_3 = 2 (1/5)^2 f_b^2 N_\star$ triples. A fraction $f_a (f_q)$ of the triples ended up with a low separation (pericentre), so the total number of strongly interacting systems is $N_x = f_x (f_b/5)^2 N_\star$ for $x={a,q}$. 

For a binary fraction $f_b=0.1$ for low mass stars, the fractions are $f_a=0.013$ and $f_q=0.072$, we conclude that  one in $\sim 1750\  (\sim 9600)$ stars experiences a collision (efficient migration) during the course of $10$ Gyrs. In the Solar neighbourhood, the thickness of the disc is $\sim d=0.4\ \rm kpc$. Assuming that stars at galactocentric dustances between $6.5-9.5$ kpc are affected by the galactic tide, and a stellar density of $0.14\ \rm pc^{-3}$ we end us with $2\times0.4\times 3 \times 3 \times 1.4\times10^9 \approx 10^9$ stars.

Similar estimates can be made on more massive stars. If the number of massive star is $f_{\rm massive} N_\star$ with $f_{\rm massive}$ a normalization parameter that depends on the mass function, $f_{\rm in}$ is the fraction of massive binaries ($M>M_\odot$) and $f_{\rm out}$ is the fraction of wider low-mass - massive binaries, then  the number of total eccentric binary interactions is $N_e = (f_a + f_e) (f_{\rm in}f_{\rm out}/25)2 f_{\rm massive} N_\star$. Here, the massive binary fraction is larger, $f_{\rm in} \approx 0.3$, while for the wider binary we conservatively assume $f_{\rm out} \sim  f_2 =0.1$. From our RG simulations, We have in this case $f_a+f_e=0.076$. The fraction of massive stars from a Kroupa IMF is \citep{kroupa2001, mic_sha21} $f_{\rm massive}\approx 0.1$, but their binary fraction is larger. Finally, we have $N_e \approx 1.8\cdot10^4$, or one of every $\sim 55000$ stars

We can further deduce the branching ratios of individual encounters from table \ref{tab:1}. Due to the uncertainty in the mass of the Galaxy and the mass function, we normalize the branching ratios of the occurence rate per $10^6$ stars, which is then easier to convert, depending of the modelling of individual galaxies.

\begin{table}
\begin{center}
\begin{tabular}{|c|c|c|c|} 
 \hline
  & MS & RG & WD \\ [0.5ex] 
 \hline\hline\
 MS & 8.57 & 3.46 & 3.3 \\ 
 \hline
 RG &  & 0.24 & 2.28 \\
 \hline
 WD &  &  & 0.48 \\ [1ex] 
 \hline
\end{tabular}
\end{center}
\caption{Occurence rate of various stellar type encounters per $10^6$ stars. }\label{tab:2}
\end{table}

\cite{kaib2014} estimated the collision rate of wide binaries in the field at the Solar neighbourhood to be $\sim 1/500$, and a collision per $10^4$ stars if $5\%$ of wide binaries are assumed. The collisions of wide binaries are dominated by flybys, while in our case the Galactic tide should have the dominant contributon \cite[see e.g. flux estimates for both in ][]{ht86}. We have a factor of six larger value. If we rescale our wide binary fraction from  $0.1 \times 0.2 = 2\%$ to $5\%$, and increase the Solar density by a factor of $2$, we will have at least an order of magniude more collisions due to wide triples. The combined effects of flybys and Galactic tide can further increase the encounter rates.

The latter numbers are derived under the assumption that the Galactic potential is roughly constant. In the inner parts of the Galaxy the density is larger, the galactic potential is stronger and the Galactic timescale is also faster, so binaries on less wide orbits, maybe $\sim 10^3 \rm au$, are significantly perturbed. Finally, the non-spherical nature of nuclear star clusters for closer triples could also be relevant \citep{petrovich17, hamers_vrr, bub2020}

The stellar field is also collisional at  smaller separations, and stochastic changes in the orbital elements may lead to more close encounters \citep{kaib2014, michaely2016, michaely2019}. On the other hand, binaries at $\sim 10^4\ \rm au$ could be unstable due to the strong tidal force, and non-vertical tides could also be important. \citep{kaib2014} estimated the rates of wide-binaries collisions for a wide range of galactocentric distances and found differences of roughly a factor of a few. Altough it's tempting to extrapolate \protect{\citep{kaib2014}}'s results to our rate estimate, we cannot assess of the change of our rates since triple evolution is more complex; a detailed study of these issues is beyond the scope of the current paper.

In the next section we briefly discuss the astrophysical implications of each type of CATGATE-catalyzed close encounters we identify.

 \subsection{Astrophysical implications} \label{implications}

 {\bf MS-MS collisions and blue stragglers:} The collisions or mass transfer of close binary MS-MS stars may lead to the formation of blue-straggler stars (BSS). BSS are found in all stellar populations, predominantly in open \citep{rain2021} and globular clusters \citep{bssgc}, but also in field stars with similar frequency to blue horizontal branch (BHB) stars \citep{santucci2015} and the Galactic halo with volumetric density of $\sim 3.4\pm0.7 \times 10^{-5} M_\odot \ \rm{pc^{-3}}$ \citep{casagrande2020}.
 
 It was already suggested that secular triple interaction can lead to BSS formation \citep{per_fab}, and collisional dynamics in the field affecting wide binaries was suggeted to give rise to MS-MS mergers \citep{kaib2014}.
 Here we estimate the importance of the CATGATE channel. In table \ref{tab:2} the number of MS-MS mergers is $\approx 8.6$ per $10^6$ stars. For low mass stars it is roughly two order of magnitude larger, or $\sim 500$ per $10^6$ stars. In order to compare it to the observes frequencies, we take the density of the stellar halo, and dividing by the local stellar density of $0.1 \ \rm pc^{-3}$ leads us to $3.4 \cdot 10^{-4}$, which is comparable to our low mass channel estimate.
 
 If we assume that every stars above $3 M_\odot$ will be a BHB star for $10\%$ of its MS lifetime, and if $\sim 1\%$ of stars are above $3M_\odot$, then $\sim 6 \times 10^{-5}$ of stars are in the BHB stage per $3M_\odot$. The observed rate of BSS exceeded the BHB stars by a factor of a few, pending on the galactocentric distance \citep{santucci2015}. This leads to a BSS rate of $\sim 5 \times 10^{-5}$ per $M_\odot$, which could be comparable to our rate in the intermediate mass channel.
 
 We conclude that the CATGATE channel could explain most of BSS stars in the field for favourable conditions. 
 
 {\bf MS-MS collisions: optical transients and red novae:} The collision or inspiral or two stars may produce a transient event such as luminous red novae (LRN) and intermediate luminous optical transients \citep{soker2006, ilot1}. Our channel may produce few transients per $10^5\ \rm yr$  in the Solar neighbourhood alone, and probably larger rate closer to the Galactic centre.
 
{\bf MS-WD encounters, cataclysmic variables and Iax SNe projenitors:} The rate of MS-WD collisions from wide binaries in the field for a Milky-Way Galaxy model was estimated in \cite{mic_sha21}, and is around $10^{-4}\ \rm{yr}^{-1}$, or around $10^6$ mergers per Galaxy of mass $\sim 10^{11}$ stars, 
which places the merger rate per $M_\odot$ to be $10^{-5}$. Our rate is roughly comparable. Most of the WD-MS encounters are inspirals which could later lead to type Ia's  \citep{michaely2021}. Collisions could also be possible especially in the non-secular regime (see sec. \ref{sec. 5.4}), where the MS will essentially be sheared apart, and an energy of $\sim 10^{48}\ \rm erg$ will be released over the course of weeks-year, with typical peak luminosity around $L=10^{7} - 10^{9}\ \rm L_\odot$ \citep{shara78}. Finally, low-mass stars might be tidally disrupted, leading to transient events \citep{per+16}. 
 
If tidal circularization is efficient, MS-WD encounters could be sources or cataclysmic-varibale (CV) stars and soft X-ray sources. The initial volume density of $\sim 10^{-4}\ \rm pc^{-3}$ \citep{pretorius2007} was recently pinned down to $4.8_{-0.8}^{+0.6} \times 10^{-6} \ \rm pc^{-3}$ owing to GAIA data \citep{gaia_cv}. For a stellar number density of $0.1\ \rm pc^{-3}$ the rate of CVs is $4.8\times 10^{-5}$ per $M_\odot$. Our estimated rate of MS-WD encounters is short by about a factor of $\sim 5$, thus the CATGATE channel could account for a significant fracion of CVs. CVs with low mass WD could merge faster, produce a transient of $\sim 10^5L_\odot$, reignite the hydrogen shell and form a giant \citep{met2021}.

{\bf RGs, sDB stars and Common envelope: } Mergers of WD with RG may lead to sDB stars. The inspiral of a WD and a RG (or MS and RG) may lead to CEE.
In particular, the CATGATE channel may provide the initial conditions for the eccentric CE channel recently studied \citep{glanz2021}.

 {\bf WD-WD collisions and type Ia supernovae:} The results of a direct WD-WD collisions is thought to be a promising channel for Type Ia Supernovae \citep{katz2011, kushnir2013, michaely2021}. In terms of stellar mass, the observed rate is around $10^{-3}$ per $M_\odot$ \citep{maoz2014, maoz17}. \cite{toonen2018} found that direct collisions of triples is a few times $10^{-7}$ per $M_{\odot}$, which accounts for at most $0.1\%$ of the observed rate. In our case, if we assume a typical mass of $1\ M_\odot$, the rate per $M_\odot$ is 
$\mathfrak{R}_{\rm Ia}  \approx 4.8\cdot 10^{-7}$,
which is comparable to \citep{toonen2018}, and equally accounts for $\lesssim 0.1\%$ of the observed rates. As we show in Sec. \ref{sec. 5.4} non-secular fluctuations may lead to direct collisions if $\log_{\rm 10}(1-e) \lesssim 10^{-4}$, which occurs at around $2\%$ of the simulated systems. Even if assuming they all occur in the double WD regime, this leads to an increase of at most one order of magnitude, so our rate could be $\lesssim 1\%$ from the observed rate.

Production of interacting mass-transfering WD binaries may also contribute to the production of type Ia supernovae through the single degenerate channel \citep{whe+73} or the production of type Ia SNe \citep{jor+12}, but binary evolution channels would be far more efficient. 
 
 {\bf Stability of planetary systems:} \cite{kaib2013} pointed out that a significant fraction of planetary systems in wide binaries may be unstable or gain significant eccentricity over Gyr timescales. Here, even binaries with separation of $\sim 100\ \rm au$ may become eccentric and render some of the planetary systems unstable. The stability of planetary systems in wide binary/triple context deserves more focused studies.  
 
 {\bf Massive stars and compact objects:} We analysed only intermidiate mass stars that form WDs. More massive stars form neutron stars (NS) and black holes (BH). The formation of a NS generally accompanied with a large natal kick, which will most likely destabilise the wide triple system, unless electron-capture SNe may produce low-velocity kicks \citep{willcox2021}. A BH may form through a direct collapse or will have significant fallback so kicks can be essentially quenched. Both cases could become unbound due to significant mass loss or form dynamically later on after their rapid stellar evolution. WD-NS/BH merger may produce distinctive transients \citep{zenati2020, bobrick2021}, while BH-BH mergers and gravitational-wave projenitors had been recognised as an important channel in the colisional field environment \citep{tbh}. CATGATE may produce compact-objects mergers, but the initial conditions of the projenitors are complex and deserve a more detailed population synthesis modelling with much more detailed stellar evolution of binary \citep{compas} and triple \citep{toonen2018, hamers21} massive stars, and considerations of the natal kicks.
 
 {\bf Wide triples properties} 
 Besides the unique transient features and final merger/inspiral product that may rise from each encounter discussed in above, several corollaries on the wide binary/triple population may drawn:
 
 {\bf Anisotropy in the GAIA sample:} Wide binaries with large inclination angle with the galactic plane become eccentric and could be unstable. Thus, constraining the orbits of wide binaries, if possible, and observing a dearth of polar orbits may pave the way to understanding the dynamical evolution of such binaries. Encounters with passing stars may randomize the inclination angles again, thus a qualitative comparison between the replenishment time and the loss time are required. The inclination distribution may be separation-dependent.
 
 {\bf Wide triples:} The current detections of wide binaries search for stars on roughly the same portion of the sky with close proper motions (the orbital velocity is negligible) and then running a data-analysis pipeline the estimates the probability of a true wide binary \citep{hartman2020}. If one of the stars is part of a tighter binary of $\sim 100\ \rm au$, then the relative velocity contaminates significantly the proper motions and the wider binary is not detected. A possible solution could be calculating the centre of mass velocity of these binaries as if they are single stars and looking for wider counterparts in the data.
 
 {\bf Super-thermal eccentricity distribution:} It had been recently deduced that the eccentricity distribution becomes super-thermal for ultra-wide binaries \citep{hwa+21}\footnote{While this paper was under review, \cite{hamilton22} shown that an initially thermal eccentricity distribution of wide binaries cannot evolve into super-thermal.}. If the zero-ago eccentricity is thermal, then dynamical evolution due do isolated triples, CATGATE may shift the eccentricity distribution to super-thermal. Wide binaries perturbed undergoing galactic tides alone will keep a thermal distribution \protect{\citep{hamilton22}}. Perhaps the combination of feedback from the inner binary, or combinations with flybys could alter the eccentricity distribution. If uncorrelated flybys were the only cause, the eccentricity distribution would have remained thermal.

 \subsection{Caveats} \label{sec. 5.4}
 
 {\bf Neglected effects and tidal dissipation:} We assumed that the widest binary is in the disc midplane and that only the vertical tide is the most relevant. Recent Gaia data allows the reconstruction of the Galactic potential \citep[and potential dark-matter density,][]{buschmann2021} in much more detail. Different spatial galactocentric distances might also change the overall dynamics and encounter rate \citep[see comparison in ][]{kaib2014}. The combined effects of realistic galactic potentials, different spatial locations and the cumulative effects of flybys could change the results, but are beyond the scope of this paper. Similar effects could domiante the collisional evolution of triple in cluster potentials \citep{hamilton1, hamilton2, hamilton3}.
 
 The exact details of tidal dissipation are also uncertain, especially in the dynamical tide regime. Nevertheless, tidal interactions are expected to be important only during a close approach. The primary focus of this paper was to understand which fraction of wide triples experience close approaches to begin with. The strength of tidal disispation will change the branching ratio of merging/inspiraling orbit, but the the overall number. Indeed, \cite{kaib2014} demonstrated that for wide binaries, the details of tidal dissipation do not alter the overall rates too much.
 
 {\bf Breakdown of double averaging:} We use a secular code with corrected double averaging \citep{luo16, gpf18, mangipudi2021}. This means that effectively the secular evolution is qualitatively similar to N-body integration as long as the total angular momentum does not vary too much. Qualitatively, (corrected) secular evolution is preserved as long as \citep[e.g.][]{ll18}
 \begin{equation}
     \sqrt{1-e^2} \gtrsim \epsilon_{\rm SA}^2,
 \end{equation}
 where $\epsilon_{\rm SA}$ is given in Eq. \ref{eq:eps_sa}. \cite{ant2014} gives a similar criterion 
 \begin{equation}
     \sqrt{1-e} \gtrsim 8\cdot 10^{-3} \frac{2m_2}{m_{\rm in}} \left(\frac{10 a_1}{a_2 (1-e_2)} \right)^3. \label{eq:n-body-crit}
 \end{equation}
 For the extreme case of $a_2 (1-e_2) = 10 a_1$ and equal masses we have $1-e\gtrsim 6.4\cdot 10^{-5}$, or $\log_{\rm 10}(1-e) \gtrsim -4.2$. This means that most of the systems are well described by secular evolution, while the small tail of highly eccentric one could be non-secular. This is potentially promising for WD-WD mergers that require extreme eccentricities, so their rates could be enhanced by non-secular evolution.
 
 Finally, as pointed out by \cite{gpf18}, most systems may become dynamically unstable before achieving condition in Eq.  \ref{eq:n-body-crit}.
 
\section{summary} \label{sec:6}

We study the influence of a Galactic tide on triple stellar systems where the outer binary is ultra wide $\gtrsim 10^4\ \rm au$. When the secular Lidov-Kozai and Galacitc tidal timescales are comparable, chaotic evolution drives the inner binary into extremely large eccentricies. Contrary to the standard ZLK case (where the triple is isolated and almost polar misalignment is required), the chaotic evolution mechanism is robust for wide range of initial conditions, provided that the secular timescales overlap. It therefore provide arobust novel channel for close interaction of stars in ultra-wide triple systems.

We simulated low-mass ($0.5-1M_\odot$) and high-mass ($1-8M_\odot$) triple systems using the secular code \href{https://github.com/eugeneg88/SecuLab}{\texttt{SecuLab}} with averaging correction, GR and equilibrium and dynamical tides, and simplified stellar evolution. Our ChAotic-Triple-GAlactic-Tide-Evolution (CATGATE) mechanism is the dominant one for close stellar encounters in the field (with Solar neighbourhood typical condtions), overtaking the collisional flybys by a factor of a few. It therefore provides a novel mechanism for producing closely interacting binaries, catcalysmic variables, and various stellar merger and collsions products and potential transeint phenomena. It would also affect the evolution of planetary systems in such wide triples.

The final outcomes and occurrence rates of close encounters depends on the stellar types involved and assumptions made on triples' population in the Galaxy. We discuss the transients and the final object formed in these encounters, as well as potential observational signatures in sec. \ref{implications}. Future studies may refine the uncertainties is our assumptions; namely, detailed stellar evolution, exact N-body intergration and tidal models, as well more realistic Galactic potential. Future observations of irregular bursts, mass-gyrochronology mismatch (such as for blue stragglers) and enhanced stellar oscillations from asteroseismology could provide evidence of eccentric encounters and/or mergers. 

Extension of the CATGATE mechanism for higher mass system producing netutron stars and black holes may also give rise to production of GW sources. The potential efficeincy of such GW soucres channel, however, would be highly dependend on the types of natal kicks given to compact objects and the survival of wide triple to secular timescales. These issues will be explored elsewhere.
\section*{Data availability statement}
The data that support the findings of this study are available from the corresponding author upon reasonable request.

\section*{Acknowledgements}
We thank Hila Glanz, Ilya Mandel and Mor Rozner for useful discussions. We thank Cristobal Petrovich for discussions and comments on the manuscript. EG and HBP acknowledge support for this project from the European Union's Horizon 2020 research and innovation program under grant agreement No 865932-ERC-SNeX.








\appendix

\section{Fraction of unstable systems}

Although we initialize stable systems, they may render unstable if the outer eccentricity exceed a certain limit. For an initial eccentricity $e_0$ and inclination $i_2$, the final eccentricity is $e_2=(1-Kx^2)^{1/2}$, where $K=5/4$ for the Galactic tide case (and $K=5/3$ for LK case) and $x=\cos i_2 (1-e_0^2)^{1/2}$. For simplicity we'll assume that $e_0\ll1$ and ignore it for now. We know that the distribution function of $x$ is uniform between $0$ and $1$, therefore the distribution function of $e_2$ is $\pi(e_2) = |dx/de_2| = e_2/(1-e_2^2)^{1/2}/K^{1/2}$. There are no oscillations if $x>K^{-1/2}$, so it also needs to accounted for via the Dirac delta-distribution. The final distribution is 

\begin{equation}
    \pi(e_2)=A\delta(e_2) + \frac{e_2}{K^{1/2}(1-e_2^2)^{1/2}}
\end{equation}
where $A=1-K^{-1/2}$ is determined to ensure the distribution is normalized.

If the system is unstable for a fixed final eccentricity $e_2=e_f$, the probability to achieve this eccentricity is \begin{equation}
    P(e_2\ge e_f) = \intop_{e_f}^1 \pi(e_2)de_2 = \sqrt{\frac{1-e_f^2}{K}} = 2\sqrt{\frac{1-e_f^2}{5}}
\end{equation}
where in the latter equality we used $K=5/4$.  

Formally, we need to account for $e_0$ and marginalize over it to get the distribution of $x$. $e_f$ also depends on $e_2$ and the semimajor axis ratio $a_1/a_2$. A proper marginalization over each of these distribution is possible but tedious, but a simple estimate can be made if we assume that on average the separation ratio is $a_1/a_2 \approx 10^{-2}$ and the binary is unstable if $a_2(1-e_f)/a_1 \lesssim 5$, which leads to $e_f=0.95$. Taking this estimate leads to a remarkable excellent fit with the simulations $\approx 0.28$ which is off by less then $0.5\%$. 

\bsp	
\label{lastpage}
\end{document}